\documentclass[%
 reprint,
 superscriptaddress,
 amsmath,amssymb,
 aps,
 pra,
]{revtex4-2}

\usepackage{physics}
\usepackage{graphicx}
\usepackage{dcolumn}
\usepackage{bm}
\usepackage[colorlinks,
            urlcolor=red,
            linkcolor=blue,
            anchorcolor=blue,
            citecolor=green
            ]{hyperref}

\usepackage{pifont}
\usepackage{subfigure}
\usepackage{balance}
\allowdisplaybreaks[3]

\newcommand{\vect}[1]{\mathbf{#1}}
\newcommand{\Eq}[1]{Eq.~\eqref{#1}}
\newcommand{\eq}[1]{\eqref{#1}}
\newcommand{\Sec}[1]{Sec.~\ref{#1}}

\newcommand{\beq}{\begin{equation}}
\newcommand{\eeq}{\end{equation}}
\newcommand{\I}{\text{i}}
\newcommand{\script}[2]{{(#1, #2)}}
\newcommand{\scriptp}[1]{{(#1)}}
\newcommand{\sh}[3]{Y_{#1}^{#2}(\hat{\vect{#3}})}
\newcommand{\shp}[3]{Y_{#1}^{#2}(\hat{\vect{#3}}_1)}
\newcommand{\diff}{\text{d}}
\newcommand{\autoparen}[1]{\left(#1\right)}
\newcommand{\autobra}[1]{\left[#1\right]}
\newcommand{\mt}{\mathcal{T}}
\newcommand{\ms}{\mathcal{S}}
\newcommand{\msp}{\mathcal{S}_1}
\newcommand{\mg}{\mathcal{G}}
\newcommand{\brho}{\bm{\rho}}
\newcommand{\cg}[6]{C^{#1,#2}_{#3,#4;#5,#6}}
\newcommand{\kroneckerdelta}[2]{\delta_{#1, #2}}

\newcommand{\rpp}{R_1}
\newcommand{\spp}{s_1}
\newcommand{\rpz}{R_{1z}}
\newcommand{\spz}{s_{1z}}
\newcommand{\rps}{R_{1s}}
\newcommand{\tphi}{\Tilde{\phi}}
\newcommand{\tf}{\Tilde{f}}
\newcommand{\tg}{\Tilde{g}}

\begin{document}

\preprint{APS/123-QED}

\title{Three-body scattering hypervolume of two-component fermions in three dimensions}

\author{Jiansen Zhang}
 \email{Jiansenzhang\_prc@stu.pku.edu.cn}
\author{Zipeng Wang}
\author{Shina Tan}%
 \email{shinatan@pku.edu.cn}
\affiliation{International Center for Quantum Materials, Peking University, Beijing 100871, China}%

\date{\today}

\begin{abstract}
We study the zero-energy collision of three fermions, two of which are in the spin-down ($\downarrow$) state and one of which is in the spin-up ($\uparrow$) state. Assuming that the two-body and the three-body interactions have a finite range, we find a parameter, $D$, called the three-body scattering hypervolume. We study the three-body wave function asymptotically when three fermions are far apart or one spin-$\uparrow$ (spin-$\downarrow$) fermion and one pair, formed by the other two fermions, are far apart, and derive three asymptotic expansions of the wave function. The three-body scattering hypervolume $D$ appears in the coefficients of such expansions at the order of $B^{-5}$, where $B=\sqrt{(s_1^2+s_2^2+s_3^2)/2}$ is the hyperradius of the triangle formed by the three fermions (we assume that the three fermions have the same mass), and $s_1,s_2,s_3$ are the sides of the triangle.
We compute the $T$-matrix element for three such fermions colliding at low energy in terms of $D$ in the absence of two-body interactions.
When the interactions are weak, we calculate $D$ approximately using the Born expansion. We also analyze the energy shift of three two-component fermions in a large periodic cube due to $D$ and generalize this result to the many-fermion system. $D$ also determines the three-body recombination rates in two-component Fermi gases, and we calculate the three-body recombination rates in terms of $D$ and the density and temperature of the gas.
\end{abstract}

\maketitle


\section{\label{sec:introduction}Introduction}
Both theoretically and experimentally, the research of the two-component Fermi system is at the center of many research fields for a long time, such as nuclear physics and quantum field theory \cite{Field1, Field2, Field3, Field4}, and ultracold atomic physics \cite{Ultracold1, Ultracold2, Ultracold3, Ultracold4}, etc.. Especially, taking advantage of the technologies for cooling atoms and molecules \cite{ExpTech1, ExpTech2, ExpTech3, ExpTech4, ExpTech5} and tuning the interactions by Feshbach resonances \cite{Feshbach1, Feshbach2, Feshbach3, Feshbach4}, people can realize the quantum degeneracy, BCS-BEC crossover \cite{BCS-BEC1, BCS-BEC2, BCS-BEC3, BCS-BEC4, BCS-BEC5}, Fermi-Hubbard model, etc. in trapped two-component Fermi gases. The experimental progresses in ultracold Fermi gases promoted a large amount of theoretical works \cite{ExpThe1, ExpThe2, ExpThe3, ExpThe4, ExpThe5}.

In the dilute ultracold quantum gases, neutral atoms collide at very low energies, and so their de Brogile wavelengths are much larger than the range of the interaction. Under this condition, the interaction can be approximately characterized by a single parameter, i.e., the $s$-wave scattering length $a_0$. (For identical fermions in the same spin state, the $s$-wave scattering is forbidden due to the Fermi statistics and the dominant effective parameter is the $p$-wave scattering volume $\Tilde{a}_1$.) 
Apart from the two-body scatterings, three-body scatterings are also common. Similarly, one can define a parameter that characterizes the low-energy three-body scattering processes. Such a parameter, which is usually called the three-body scattering hypervolume and denoted by $D$, was first defined for the three-body scattering of identical bosons in three dimensions (3D) \cite{Tan2008}. $D$ enters the expansion of the three-body scattering wave function $\Psi$ as a parameter \cite{Tan2008}. For three identical bosons,
the most important wave function for zero-energy collision has the form \cite{Tan2008}
\begin{align}
    \Psi_\text{3-boson} &= 1 - \frac{\sqrt{3}D_\text{3-boson}}{8\pi^3B^4} + A_2 + O(B^{-5}\ln{B})
\end{align}
at large $B$, where $B=\sqrt{(s_1^2+s_2^2+s_3^3)/2}$ is the hyperradius of the triangle formed by the three particles, $s_i$ is the distance between particles $j$ and $k$ (for $i=1$, we define $j=2$ and $k=3$; for $i=2$, we define $j=3$ and $k=1$; for $i=3$, we define $j=1$ and $k=2$), and $A_2$ is a part of the expansion that depends on the two-body scattering parameters and $s_1,s_2,s_3$ only.

The very definition of the three-body scattering hypervolume depends on the statistics of the particles and the spatial dimensionality.
Up to now, the three-body scattering hypervolumes have been defined for identical bosons \cite{Tan2008}, for distinguishable particles \cite{Wang2021-1}, for two identical bosons and a particle with a different mass \cite{mestrom1}, for identical spin-1 bosons \cite{mestrom2}, for identical spin-polarized fermions in 3D \cite{Wang2021-2}, in two dimensions (2D) \cite{Wang2022} and in one dimension (1D) \cite{Wang2023}, and for identical bosons in 2D \cite{Liang2024}. However, studies of the three-body scattering hypervolume in two-component fermionic systems, with at least two fermions in different spin states, are notably absent. So we study the three-body scattering hypervolume in the two-component fermionic system in this manuscript.

The three-body scattering hypervolumes can be regarded as three-body analogs of the two-body $s$-wave scattering length $a_0$, and they are fundamental parameters that determine the effective strengths of three-body interactions with small collision energies. 
The three-body scattering hypervolumes determine the effective three-body coupling constants in the effective field theory\cite{Tan2008, effectivefield}.
The three-body scattering hypervolumes affect essentially \emph{all} the $N$-body ($N\ge 3$) and
 many-body properties of the system, such as the three-body $T$-matrix elements \cite{Tan2008},
 the energies of dilute many-body systems \cite{Tan2008}, and the three-body recombination rates \cite{Wang2021-1, Wang2021-2, Wang2022, Wang2023, Liang2024}.


In order to theoretically understand ultracold quantum gases with short-range interactions (having a finite interaction range $r_e$), one can
define the few-body wave functions for zero-energy collisions in the center-of-mass (COM) frame and use them as some of the building blocks for understanding the many-body wave functions. There are infinitely many such two-body wave functions with zero collision energy, but the most important one of them is the wave function which grows at the \emph{slowest rate} when the two-body distance $s\to\infty$ (because the faster the wave function grows at large $s$, the less likely it is for the two particles to be found within the range of interaction). For two-component fermionic systems, such wave function is the $s$-wave collision wave function, and it has the following formula at $s>r_e$:
\beq
    \phi(\vect s)=1-\frac{a_0}{s},
\eeq
where $a_0$ is the $s$-wave scattering length between two fermions in different spin states.
$\phi(\vect s)$ depends on the details of the two-body interaction at $s<r_e$.
Note that $\phi(\vect s)$ scales like $s^0$ at large $s$.
In comparison, the $p$-wave and the high-partial-wave two-body wave functions for zero collision energies
grow like $s^l$ with $l\ge1$ at large $s$, and are usually less important than the $s$-wave collision wave function $\phi(\vect s)$
for purposes of understanding the two-body interaction effects.

Similarly, we have infinitely many three-body wave functions for zero collision energies in the COM frame,
but the most important one of them (for purposes of understanding the three-body effective interaction effects in ultracold quantum gases) is usually the one
which grows at the \emph{slowest rate} when the three-body hyperradius $B\to\infty$. 

For the three-body system containing two spin-$\downarrow$ fermions and one spin-$\uparrow$ fermion,
the zero-energy three-body wave function can \emph{not} scale like $B^0$ at $B\to\infty$ (because
this would violate Fermi statistics), and so the most important zero-energy three-body wave functions scale like $B^1$
at $B\to\infty$. In the COM frame, there are only three such wave functions that are linearly independent, and they have orbital angular momentum quantum number $L=1$ and magnetic quantum numbers $M=1,0,-1$. So in this paper, we study the wave function for three two-component fermions, two of which are in the spin-$\downarrow$ state and one of which is in the spin-$\uparrow$ state, with zero collision energy, with $L=1$ and $M=0$, and with asymptotic behavior
$$
\Psi\sim B^1,~~B\to\infty.
$$
(The wave functions with $L=1$ and $M=\pm1$ can be obtained by applying the operators $J_\pm$ on the wave function with $L=1$ and $M=0$. Here $J_\pm$ is the ladder operator defined in Appendix \ref{app:expansion}.)
We find three asymptotic behaviors of the three-body wave function: one 111-expansion and two 21-expansions. The 111-expansion is the asymptotic expansion of the three-body wave function when the three pairwise distances $s_1,s_2,s_3$ go to infinity simultaneously, the first kind 21-expansion is the asymptotic expansion when two spin-$\downarrow$ fermions are kept at a fixed distance and the spin-$\uparrow$ fermion is far away from the two, and the second kind 21-expansion is the asymptotic expansion when one spin-$\uparrow$ fermion and one spin-$\downarrow$ fermion are kept at a fixed distance and another spin-$\downarrow$ fermion is far away from the two. The three-body scattering hypervolume $D$ (which is independent from the two-body parameters) appears at the order of $B^{-5}$ in the 111-expansion, and this implies that the dimension of $D$ is length to the \textit{sixth} power.

In \Sec{sec:asymptotics},  
we show the resultant 111-expansion and 21-expansions, which arise from formally solving the three-body Schr\"{o}dinger equation at large $B$ asymptotically. When the two-body interactions support bound states, we show that $D$ acquires a negative imaginary part related to the probability amplitudes of the three-body recombination processes.
In \Sec{sec:3bodyscattering}, we show that the parameter $D$ determines the three-body $T$-matrix element for low energy three-body collisions in the absence of two-body interactions.
In \Sec{sec:born_approximation}, we study the three-body wave function using Born approximation when the interaction potentials are weak. We derive an approximate formula for $D$ in terms of the weak potentials. In \Sec{sec:energy_shift}, we consider three two-component fermions in a large periodic cubic box and calculate the shift of the energy due to a nonzero $D$. We then generalize the result to the $N$-body system, where $N=N_\uparrow+N_\downarrow$, and $N_\sigma$ is the number of spin-$\sigma$ fermions for $\sigma=\uparrow, \downarrow$. In \Sec{sec:three-body_recomb}, we suppose that the interaction potentials can support two-body bound states and calculate the three-body recombination rates of two-component Fermi gases in terms of the
three-body scattering hypervolumes.

\section{\label{sec:asymptotics}Asymptotic behavior of the three-body wave function}
We consider three two-component fermions with the same mass $m_F$, and two of them are in the spin-$\downarrow$ state and labeled as particles 1 and 2, one of them is in the spin-$\uparrow$ state, and labeled as particle 3. 
Let $\vect r_i$ be the position vector of the $i$th particle.
We also define the Jacobi coordinates $\vect s_i$ and $\vect R_i$, the hyperradius $B$, and the hyperangles $\Theta_i$ \cite{Jacobi1, Jacobi2}:
\begin{subequations}
\begin{align}
    \vect s_i &\equiv\,\vect r_j-\vect r_k,\label{eq:sidef} \\
    \vect R_i &\equiv\,\vect r_i-(\vect r_j+\vect r_k)/2,\label{eq:ridef} \\
    B &\equiv\,\sqrt{(s_1^2+s_2^2+s_3^2)/2}=\sqrt{R_i^2+\frac{3}{4}s_i^2}, \\
    \Theta_i &\equiv\, 
    \arctan\frac{2R_i}{\sqrt{3}s_i},
\end{align}
\end{subequations}
where ($i, j, k$) is any \emph{even} permutation of $(1, 2, 3)$. 
One can show that $s_i=\frac{2}{\sqrt{3}}B\cos\Theta_i$, and $R_i=B\sin\Theta_i$.
For later convenience, we define short-hand notations
\begin{subequations}
\begin{align}
\vect s&\equiv\vect s_3,\label{eq:sdef} \\
\vect R&\equiv\vect R_3,\\
\Theta&\equiv\Theta_3.
\end{align}
\end{subequations}

In the center-of-mass (COM) reference frame, the three-body zero-energy scattering wave function $\Psi(\vect r_1,\vect r_2,\vect r_3)$ satisfies the Schr\"{o}dinger equation
\beq\label{eq:3bodyschrodinger}
    \sum_{i=1}^3 \left[-\frac{\hbar^2}{2 m_F} \nabla_i^2 + V_i(s_i)\right] \Psi + U(s_1, s_2, s_3) \Psi=0,
\eeq
where $\hbar$ is the reduced Plank constant,
$V_i(s_i)$ is the two-body interaction potential between the $j$th and the $k$th particles and $V_1=V_2$, and $U(s_1,s_2,s_3)$ is the three-body interaction potential. We assume that all the interaction potentials are invariant under overall translations, rotations, and finite-ranged [when $s_i>r_e$, $V_i(s_i)=0$; when $s_1>r_e$ or $s_2>r_e$ or $s_3>r_e$, $U(s_1,s_2,s_3)=0$].
Since we study the problem in the COM frame, the wave function is translationally invariant, i.e.,
\beq\label{eq:translationinvariant}
    \Psi(\vect r_1,\vect r_2,\vect r_3)=\Psi(\vect r_1+\delta\vect r,\vect r_2+\delta\vect r,\vect r_3+\delta\vect r)
\eeq
for any displacement $\delta\vect r$. Also, $\Psi(\vect r_1,\vect r_2,\vect r_3)$ should be antisymmetric under the exchange of particles 1 and 2, because they are identical fermions in the same spin state.

To uniquely determine the wave function $\Psi$, we also need to specify the asymptotic behavior of $\Psi$ when all three particles are far apart from each other. Let $\Psi_0$ be the leading order term in the expansion of $\Psi$ when $s_1$, $s_2$, and $s_3$ approach infinity simultaneously. We assume that $\Psi_0$ scales as $B^p$ at large $B$. $\Psi_0$ should satisfy the Laplace equation $(\nabla_1^2+\nabla_2^2+\nabla_3^2)\Psi_0=0$, so $\Psi_0$ is a harmonic polynomial of degree $p$. For dilute ultracold gases,
the most important scattering channel requires the minimum value of $p$,  because the smaller $p$ is, the more likely for the three particles to come into the range of interaction. When $p=0$, $\Psi_0$ is just a constant, which violates the antisymmetry of the wave function under the exchange of particles 1 and 2. Thus, we find that the minimum value of $p$ for this system is $p_{\text{min}}=1$, and the $\Psi_0$ for this minimum value of $p$ can take three linearly independent forms: $s_x$, $s_y$, and $s_z$. We will first focus on the three-body wave function with
\begin{equation}\label{Psi0}
    \Psi_0=s_z,
\end{equation}
This leading order term has total orbital angular momentum quantum number $L=1$ and magnetic quantum number $M=0$. There are two other linearly independent leading order terms that also scale as $B^1$ and they can be written as $-(s_x+ \I s_y)/\sqrt{2}$ and $(s_x-\I s_y)/\sqrt2$. These three leading order wave functions can be distinguished by the magnetic quantum number $M=0, \pm1$ along the $z$ direction. In the following, we will study the wave function for the state $\ket{L=1, M=0}$. 
In Appendix~\ref{app:expansion} we show how to generate the expansions for the wave functions of the states
$\ket{L=1, M=1}$ and $\ket{L=1, M=-1}$.

\subsection{The 111-expansion and the 21-expansions}
When three particles are far apart from each other, which means that the pairwise inter-particle distances $s_1$, $s_2$, $s_3$ all go to infinity for any fixed ratio $s_1:s_2:s_3$, one can expand $\Psi$ in powers of $B^{-1}$:
\begin{equation}\label{eq:111def}
    \Psi=\sum_{p=-1}^\infty\mt^{(-p)}(\vect r_1, \vect r_2, \vect r_3),
\end{equation}
where $\mt^{(-p)}$ scales as $B^{-p}$. Equation~\eqref{eq:111def} is called the 111-expansion. 
When particle 1 and particle 2 are kept at a fixed distance $s$ and particle 3 is far away from them, one can expand $\Psi$ in powers of $R^{-1}$:
\begin{equation}\label{eq:1st21def}
    \Psi=\sum_{q=0}^\infty\ms_3^{(-q)}(\vect R, \vect s),
\end{equation}
where $\ms_3^{(-q)}$ scales as $R^{-q}$, and this expansion is called the first kind 21-expansion. 
When particle 2 and particle 3 are kept at a fixed distance $s_1$ and particle 1 is far away from them, one can expand $\Psi$ in powers of $R_1^{-1}$:
\begin{equation}\label{eq:2nd21def}
    \Psi=\sum_{q=-1}^\infty\msp^{(-q)}(\vect R_1, \vect s_1),
\end{equation}
where $\msp^{(-q)}$ scales as $R_1^{-q}$, and this expansion is called the second kind 21-expansion.

The procedure for determining the 111-expansion and the 21-expansions is shown in Appendix~\ref{app:expansion}. The resultant 111-expansion is
\begin{widetext}
\begin{align}\label{eq:111expansion}
    \Psi =& s_z\left(1 -\frac{3\sqrt{3}D}{8\pi^3B^6}\right) -\frac{a_0R_{1z}}{s_1}+\frac{a_0R_{2z}}{s_2}  \nonumber \\
    &-\frac{a^2_0}{\pi}\left[\frac{R_{1z}}{s_1R_1}(\cot{\Theta_1}-\Theta_1(1+\cot^2\Theta_1))-\frac{R_{2z}}{s_2R_2}(\cot{\Theta_2}-\Theta_2(1+\cot^2\Theta_2))\right] \nonumber \\
    &-\frac{3\Tilde{a}_1s_z}{s^3}+\frac{3a_1}{2}\left(\frac{s_{1z}}{s_1^3}+\frac{s_{2z}}{s_2^3}\right)+\frac{2\omega a^3_0}{\pi}\left[\frac{R_{1z}}{s_1R_1^2}(1-\Theta_1\cot{\Theta_1})-\frac{R_{2z}}{s_2R_2^2}(1-\Theta_2\cot{\Theta_2})\right] \nonumber\\
    & +\frac{2a_0\omega_3}{\pi}
    \left[\frac{R_{1z}}{R_1^3s_1}\left(\Theta_1-\frac{1}{2}\sin2\Theta_1\right)-\frac{R_{2z}}{R_2^3s_2}\left(\Theta_2-\frac{1}{2}\sin2\Theta_2\right)\right] \nonumber \\
    & + \sum_{i=1}^3 \frac{\beta_{-4,i}}{R_i^2s_i^3}\autobra{s_{iz}\sin^2\Theta_i(-12\omega\cos^2\Theta_i+9\sqrt{3}-6\pi)+\frac{2\omega'}{R_i^2}R_{iz}\mathbf{R}_i\cdot\mathbf{s}_i\sin^4\Theta_i} -a_0\omega_4\left(\frac{R_{1z}}{B^4s_1}-\frac{R_{2z}}{B^4s_2}\right) \nonumber \\
    & + \frac{2a_0\Omega_5}{9\pi}\sum_{i=1}^2 (-1)^{i-1}\frac{R_{iz}}{R_i^5s_i}\left[24\sin^5\Theta_i\cos\Theta_i\ln\frac{b}{R_i}+\eta_{-4}(\Theta_i)\right] \nonumber \\
    & + \sum_{i=1}^2 (-1)^{i-1}\frac{9a_0a_2}{R_i^2s_i^3}\left[\frac{2}{3}\eta_{-2}\autoparen{\Theta_i}\sum_{m=-2}^2\cg{1}{0}{3}{m}{2}{-m}Y_3^{m}(\hat{\vect{R}}_i)Y_2^{-m}(\hat{\vect{s}}_i) + 2\sqrt{6}\Tilde{\eta}_{-2}(\Theta_i)\sum_{m=-1}^1 \cg{1}{0}{1}{m}{2}{-m}Y_1^{m}(\hat{\vect{R}}_i)Y_2^{-m}(\hat{\vect{s}}_i)\right]
    \nonumber \\
    & +\sum_{i=1}^3\frac{\beta_{-5,i}}{R_i^3s_i^3} \Bigg\{s_{iz}\left[256\omega a_0^3\sin^3\Theta_i\cos^3\Theta_i\ln\frac{b}{R_i}+\ell_i(\Theta_i-\frac{1}{4}\sin4\Theta_i)+16\omega a_0^3\eta_{-3}(\Theta_i)\right] -\frac{3R_{iz}}{R_i^2}(\vect R_i\cdot\vect s_i)\ell_i\,\Tilde{\eta}_{-3}(\Theta_i)\Bigg\} \nonumber \\
    &+O(B^{-6}\ln^n B),
\end{align}
where $D$ is the three-body scattering hypervolume, which appears at the order of $B^{-5}$ and has the dimension of length to the \emph{sixth} power, $Y_l^m$ is the spherical harmonics \cite{SH},  
$n$ is a nonnegative integer, $\cg{J}{M}{l_1}{m_1}{l_2}{m_2}=\braket{l_1,m_1;l_2,m_2}{J,M}$ is the Clebsch-Gordan coefficient,
and
\begin{align}
    \omega =&\, \sqrt{3}-\pi/3, \\
    \omega' =&\, -15\sqrt{3}+8\pi,\\
    \omega_3 =&\, 3\Tilde{a}_1+3a_1/2+(7\sqrt{3}/6-4/\pi)\omega a^3_0-3a_0^2r_0/8, \\
    \omega_4 =&\, \autobra{3\omega a_0^3r_0/2-4\omega\omega_3a_0+\autoparen{15\sqrt{3}-2\pi}\autoparen{a_0\Tilde{a}_1+a_0a_1/2}}/\pi, \\
    \Omega_5 =&\, \frac{a_0^2}{2\pi}\autoparen{-15\sqrt{3}+2\pi}\autoparen{\Tilde{a}_1+a_1/2}-a_0\omega_4/2, \\
    \beta_{-3,1} =&\, \beta_{-3,2}= -3a_0a_1/2\pi, \\
    \beta_{-3,3}=&\, 3a_0\Tilde{a}_1/\pi,\\
    \beta_{-4,1} =&\, \beta_{-4,2}= -a_0^2a_1/2\pi,\\
    \beta_{-4,3} =&\, a_0^2\Tilde{a}_1/\pi,\\
    \beta_{-5,1} =&\, \beta_{-5,2}= -a_1/8\pi^2,\\
    \beta_{-5,3} =&\, \Tilde{a}_1/4\pi^2,\\
    \ell_1=\ell_2 =&\, 72\pi \Tilde{a}_1+9\pi (-4+a_0r_1)a_1+16\omega a_0^3(-16+3\sqrt{3}\pi),\\
    \ell_3 =&\, 36\pi a_1+9\pi a_0\Tilde{a}_1\Tilde{r}_1+16\omega a_0^3(-16+3\sqrt{3}\pi),
\end{align}
\begin{eqnarray}
    \Tilde{\eta}_{-2}(\Theta) &=& \frac{1}{12}\autoparen{-12\Theta-3\sin2\Theta+3\sin4\Theta+\sin6\Theta}, \\
    \eta_{-2}(\Theta) &=& \frac{1}{12}\left[30\cot\Theta+6\Theta(8-5\csc^2\Theta)-13\sin2\Theta-2\sin4\Theta+\sin6\Theta\right], \\
    \Tilde{\eta}_{-3}(\Theta) &=& \frac{1}{12}\autoparen{12\Theta-3\sin2\Theta-3\sin4\Theta+\sin6\Theta}, \\
    \eta_{-3}(\Theta) &=& \sin^2\Theta\autobra{2\Theta\autoparen{2\cos2\Theta+\cos4\Theta}+5\sin2\Theta+3\sin4\Theta+16\sin\Theta\cos^3\Theta\autoparen{-\gamma+\ln\sin\Theta}}, \\
    \eta_{-4}(\Theta) &=& 24\sin^5\Theta\cos\Theta\ln\sin\Theta-\sin^3\Theta(2\cos\Theta+\cos3\Theta)+3\Theta\sin^2\Theta(2\cos2\Theta-\cos4\Theta).
\end{eqnarray}
$\gamma=0.57721566\cdots$ is Euler's constant, $a_l$, $r_l$, $\tilde{a}_l$ and $\tilde{r}_l$ ($l=0, 1, 2, \dots$) are the two-body scattering parameters defined in Appendix~\ref{app:twobodyfunctions},
and $b>0$ is an arbitrary length scale upon which $D$ usually depends, namely $D=D_b$. If one changes $b$, one should also change $D$ in such a way
that the three-body wave function is not affected:
\begin{align}
    D_b-D_{b'}=\frac{8\pi}{9}a_0^3\autobra{4\omega^2\autoparen{7\sqrt{3}-24/\pi}a_0^3-18\omega a_0^2r_0-6\autoparen{3\sqrt{3}+2\pi}\autoparen{2\Tilde{a}_1+a_1}}\ln{\frac{b}{b'}}.
\end{align}
If $a_0\ne0$, $\mt^{(-5)}$ contains terms that depend on $B$ like $B^{-5}\ln B$ for fixed
hyperangles. But if $a_0=0$, such terms vanish and $D$ does \emph{not} depend on $b$. 
To understand why we have a three-body parameter in the expansion of the wave function at large $B$,
we may first consider a theoretical system with no two-body interaction and only a short-range three-body interaction,
so that the three-body Schr\"{o}dinger equation is simplified as $(\nabla_1^2+\nabla_2^2+\nabla_3^2)\Psi=0$ at $B>r_e$ (where $r_e$
is the range of the three-body interaction). Solving this equation (which may be simplified as a second order linear differential equation in the hyperradial direction for any specified hyperspherical harmonic), we find two linearly independent solutions for $L=1$ and $M=0$: $s_z$ and $s_z B^{-6}$, so the actual three-body wave function should contain a linear combination of the two at $B>r_e$, and the parameter $D$ is the coefficient in such a linear combination and it has the dimension of length to the sixth power. $D$ depends on the details of the three-body interaction in the absence of two-body interactions.
If there \emph{are} two-body interactions (as we assume in this section), the 111-expansion becomes much more complicated as shown in \Eq{eq:111expansion},
but the three-body parameter $D$ survives in such an expansion, and now $D$ depends on both the two-body interactions and the three-body interaction.

Since the leading order term $\mt^\scriptp{1}=s_z$ is antisymmetric under the exchange of particles 1 and 2, by following the zigzag procedure of determining the 111-expansion and the 21-expansions, shown in Appendix~\ref{app:expansion}, we have verified that
these expansions are antisymmetric (with respect to particles 1 and 2) at each order automatically.

The first kind 21-expansion reads
\begin{align}\label{eq:1st21expansion}
    \Psi =&\, c^0_{1,0}\tphi^{(1,0)}(\mathbf{s}) + \sum_{m=-1}^1 c^{-1}_{1,m}\tphi^{(1,m)}(\mathbf{s}) + \sum_{m=-1}^1 c^{-2}_{1,m}\tphi^{(1,m)}(\mathbf{s}) + \sum_{m=-1}^1 \Tilde{c}^{-1}_{1,m}\tf^{(1,m)}(\mathbf{s}) + \sum_{l=1,3}\sum_{m=-l}^l c^{-3}_{l,m}\tphi^{(l,m)}(\mathbf{s}) \nonumber \\
    & + \sum_{m=-1}^1 \Tilde{c}^{-2}_{1,m}\tf^{(1,m)}(\mathbf{s}) + \sum_{l=1,3}\sum_{m=-l}^l c^{-4}_{l,m}\tphi^{(l,m)}(\mathbf{s}) + \sum_{l=1,3}\sum_{m=-l}^l \Tilde{c}^{-3}_{l,m}\tf^{(l,m)}(\mathbf{s}) 
    + \sum_{l=1,3,5}\sum_{m=-l}^l c^{-5}_{l,m}\tphi^{(l,m)}(\mathbf{s}) \nonumber \\
    &+O(R^{-6}\ln^{n_3} R),
\end{align}
where $n_3$ is a nonnegative integer, the two-body special functions $\tphi^{(l,m)}(\mathbf{s})$ and $\tf^{(l,m)}(\mathbf{s})$ are defined in Appendix~\ref{app:twobodyfunctions}, and
\begin{subequations}\label{eq:c010etc}
\begin{align}
    c^0_{1,0} =&\, 3, \\
    c^{-1}_{1,m} =&\, \sqrt{2\pi}\frac{a_0}{R}C^{1,0}_{1,m; 2,-m}Y_2^{-m}(\hat{\vect R}) -\kroneckerdelta{m}{0}10\sqrt{\pi}\frac{a_0}{R}\cg{1}{0}{1}{0}{0}{0}\sh{0}{0}{R}, \label{eq:c-11m} \\
    c^{-2}_{1,m} =&\, \frac{4\sqrt{2}\omega' a_0^2}{3\sqrt{\pi}R^2}\cg{1}{0}{1}{m}{2}{-m}\sh{2}{-m}{R} + \kroneckerdelta{m}{0}\frac{2\omega'' a_0^2}{3\sqrt{\pi}R^2}\cg{1}{0}{1}{0}{0}{0}\sh{0}{0}{R}, \\
    \omega'' =&\, 3\sqrt{3}+2\pi,
\end{align}
\end{subequations}
where $\kroneckerdelta{i}{j}$ is the Kronecker delta symbol.
The coefficients $c^{-i}_{l,m}$ and $\Tilde{c}^{-i}_{l,m}$ are functions of $\vect R$, and they satisfy
\begin{equation}
\Tilde{c}^{-i}_{l,m}=\frac{3}{4}\nabla^2_{\vect R}c^{-i}_{l,m}.
\end{equation}
When $R\rightarrow\infty$, $c^{-i}_{l,m}$ scales as $1/R^{i}$ and $\Tilde{c}^{-i}_{l,m}$ scales as $1/R^{i+2}$. The coefficients $c^{-i}_{l,m}$ are shown in Eqs.~\eqref{eq:c010etc} and Eqs.~(\ref{eq:c-start}--\ref{eq:c-end}).

The second kind 21-expansion reads
\begin{align}\label{eq:2nd21expansion}
    \Psi &= d^1_{0,0}\phi^{(0,0)}(\mathbf{s}_1) + d^0_{0,0}\phi^{(0,0)}(\mathbf{s}_1)+d^0_{1,0}\phi^{(1,0)}(\mathbf{s}_1) + d^{-1}_{0,0}\phi^{(0,0)}(\mathbf{s}_1) + \sum_{m=-1}^1 d^{-1}_{1,m}\phi^{(1,m)}(\mathbf{s}_1) \nonumber \\
    &+ \Tilde{d}^0_{0,0}f^{(0,0)}(\mathbf{s}_1) + \sum_{l=0}^2\sum_{m=-l}^l d^{-2}_{l,m}\phi^{(l,m)}(\mathbf{s}_1) + \sum_{l=0}^1\sum_{m=-l}^l \Tilde{d}^{-1}_{l,m}f^{(l,m)}(\mathbf{s}_1) + \sum_{l=0}^3\sum_{m=-l}^l d^{-3}_{l,m}\phi^{(l,m)}(\mathbf{s}_1) \nonumber \\
    & +  \sum_{l=0}^2\sum_{m=-l}^l \Tilde{d}^{-2}_{l,m}f^{(l,m)}(\mathbf{s}_1) + \sum_{l=0}^4\sum_{m=-l}^l d^{-4}_{l,m}\phi^{(l,m)}(\mathbf{s}_1) + \sum_{l=0}^3\sum_{m=-l}^l \Tilde{d}^{-3}_{l,m}f^{(l,m)}(\mathbf{s}_1) + \sum_{l=0}^5\sum_{m=-l}^l d^{-5}_{l,m}\phi^{(l,m)}(\mathbf{s}_1) \nonumber \\
    & + \Tilde{\Tilde{d}}^{-1}_{0,0}g^{(0,0)}(\mathbf{s}_1) + O(R_1^{-6}\ln^{n_1}R_1),
\end{align}
where $n_1$ is a nonnegative integer, the two-body special functions $\phi^{(l,m)}(\mathbf{s}_1)$ , $f^{(l,m)}(\mathbf{s}_1)$ and $g^{(l,m)}(\mathbf{s}_1)$ are defined in Appendix~\ref{app:twobodyfunctions}, and
\begin{subequations}\label{eq:d100etc}
\begin{align}
    d^1_{0,0} =&\, R_{1z}, \\
    d^0_{0,0} =&\, -\frac{1}{2}a_0\frac{R_{1z}}{R_1}, \\
    d^0_{1,0} =&\, -\frac{3}{2}, \\
    d^{-1}_{0,0} =&\, -\frac{2\omega a^2_0}{\pi}\frac{R_{1z}}{R_1^2}, \\
    d^{-1}_{1,m} =&\, -\frac{\sqrt{2\pi}a_0}{2R_1}\cg{1}{0}{1}{m}{2}{-m}\shp{2}{-m}{R}+\kroneckerdelta{m}{0}\frac{5\sqrt{\pi}a_0}{R_1}\cg{1}{0}{1}{0}{0}{0}\shp{0}{0}{R}.
\end{align}\end{subequations}\end{widetext}
The coefficients $d^{-i}_{l,m}$, $\Tilde{d}^{-i}_{l,m}$ and $\Tilde{\Tilde{d}}^{-i}_{l,m}$ are functions of $\vect R_1$, and they satisfy 
\begin{align}
    \Tilde{d}^{-i}_{l,m}&= \frac{3}{4}\nabla^2_{\vect R_1}d^{-i}_{l,m},\\
    \Tilde{\Tilde{d}}^{-i}_{l,m}&= \frac{3}{4}\nabla^2_{\vect R_1}\Tilde{d}^{-i}_{l,m}.
\end{align}
When $R_1\rightarrow\infty$, $d^{-i}_{l,m}$ scales as $1/R_1^{i}$, $\Tilde{d}^{-i}_{l,m}$ scales as $1/R_1^{i+2}$, and $\Tilde{\Tilde{d}}^{-i}_{l,m}$ scales as $1/R_1^{i+4}$.
The coefficients $d^{-i}_{l,m}$ are shown in Eqs.~\eqref{eq:d100etc} and Eqs.~(\ref{eq:d-start}--\ref{eq:d-end}). In \Eq{eq:2nd21expansion}, note that the coefficient $d^{-5}_{0, 0}$
depends on $D$, as shown in \Eq{eq:d-500}.

Note that the 21-expansions \Eq{eq:1st21expansion} and \Eq{eq:2nd21expansion} are applicable only when the interactions do not support any two-body bound states. If the interactions can support one or more two-body bound states, then three-body recombination will occur. In this case, two fermions may form a bound state and release the binding energy, and the bound pair and the free fermion fly apart from each other with total kinetic energy equal to the released binding energy. Thus the 21-expansions should be modified as \cite{Zhu2017}
\begin{equation}
    \Psi = \widetilde{\Psi} + \sum_{i=1}^3
    \sum_{\nu, l, l', m}c_{ill'\nu}\cg{1}{0}{l}{m}{l'}{-m}\varphi^\scriptp{m}_{ill'\nu}(\mathbf{s}_i,\mathbf{R}_i),
\end{equation}
where $\widetilde{\Psi}$ is equal to the right hand side of \Eq{eq:1st21expansion} when $\vect s$ is fixed and $R$ is large, and equal to the right hand side of \Eq{eq:2nd21expansion} when $\vect s_1$ is fixed and $R_1$ is large.
$\varphi^\scriptp{m}_{ill'\nu}(\mathbf{s}_i,\mathrm{R}_i)$ specifies that a bound pair and the remaining fermion
fly apart from each other, where $l$, $m$, and $\nu$ are the orbital angular momentum quantum number, the magnetic quantum number and the vibrational quantum number of the bound pair formed by particles $j$ and $k$, respectively, and $l'$ is the orbital angular momentum quantum number for the relative motion of the bound pair and particle $i$. $c_{ill'\nu}$ is a  coefficient. If $|l-l'|\ge2$ or $l=l'=0$, we define $c_{ill'\nu}=0$. We find
\begin{align}
    \varphi^\scriptp{m}_{ill'\nu}(\mathbf{s}_i,\mathbf{R}_i) =&\, u_{il\nu}(s_i)Y_{l}^{m}(\hat{\mathbf{s}}_i)\nonumber \\
    & \times \I^{l'+1}h^\scriptp{1}_{l'}\autoparen{\frac{2}{\sqrt{3}}\kappa_{il\nu}\mathrm{R}_i}Y_{l'}^{-m}(\hat{\mathbf{R}}_i),
\end{align}
where $h^\scriptp{1}_l$ is the spherical Hankel function of the first kind, $\kappa_{il\nu}>0$ 
is the binding wave number such that the bound pair has binding energy $\hbar^2\kappa^2_{il\nu}/m_F$, and $u_{il\nu}(s_i)$ is the radial part of the wave function for the bound pair and satisfies the Schr\"{o}dinger equation and the normalization condition
\begin{align}
    \left[-\frac{\hbar^2}{m_F}\nabla_{\mathrm{s}_i}^2+V_i(s_i)+\frac{\hbar^2}{m_F}\kappa^2_{il\nu}\right]u_{il\nu}(s_i)Y_l^m(\hat{\mathbf{s}}_i)=\,0,
\end{align}
\begin{equation}
\int_0^\infty s^2u_{il\nu}^*(s)u_{il\nu'}(s)\diff s=\frac{\delta_{\nu\nu'}}{\kappa_{il\nu}^5}.
\end{equation}
Because of the restrictions from the Fermi statistics and the parity of the leading order term, $l$ must be odd when $i=3$, and $c_{1ll'\nu}=(-1)^{l+1}c_{2ll'\nu}$ and $u_{1l\nu}=u_{2l\nu}$. 

Since the outgoing wave contributes a positive probability flux toward the outside of a large hypersurface (defined as the surface with a constant and large value of $B$ in the three-body configuration space), $D$ must gain a negative imaginary part to satisfy the conservation of probability \cite{Zhu2017, Petrov2019}. From this conservation of probability, we derive
\begin{equation}\label{eq:ImD}
    \mathrm{Im}(D)=-\frac{3\sqrt{3}}{8}\sum_{ill'\nu}\frac{|c_{ill'\nu}|^2}{\kappa^6_{il\nu}},
\end{equation}
where $\mathrm{Im}(D)$ is the imaginary part of $D$, and the summation is over all the dimer states. In Sec. \ref{sec:three-body_recomb}, we will discover the relation between $\mathrm{Im}(D)$ and the three-body recombination rate.

\section{\label{sec:3bodyscattering}Three-body low-energy scattering}
The three-body scattering hypervolume $D$ affects the $T$-matrix elements of three two-component fermions colliding at low energy.
Consider three such fermions colliding in the COM frame, assuming that particles 1 and 2 are in the spin-$\downarrow$ state with incoming momenta $\hbar\vect k_1$ and $\hbar\vect k_2$, and that particle 3 is in the spin-$\uparrow$ state with incoming momentum $\hbar\vect k_3=-\hbar(\vect k_1+\vect k_2)$. 
The particles have energy $E=\hbar^2(k_1^2+k_2^2+k_3^2)/2m_F$. 
Let $\widetilde{\Psi}^{(E)}_{\vect{q}_1\vect{q}_2\vect{q}_3}$ be the momentum-space wave function describing the scattering process, where $\hbar\vect{q}_i$ is the momentum of particle $i$, satisfying $\vect{q}_1+\vect{q}_2+\vect{q}_3\equiv 0$.
The real-space wave function for the process is the Fourier transform of $\widetilde{\Psi}^{(E)}_{\vect{q}_1\vect{q}_2\vect{q}_3}$:
\beq
\Psi^{(E)}(\vect r_1,\vect r_2,\vect r_3)=\int\frac{d^3q_1}{(2\pi)^3}\frac{d^3q_2}{(2\pi)^3}\widetilde{\Psi}^{(E)}_{\vect q_1\vect q_2\vect q_3}e^{\sum_{i=1}^3\I\vect q_i\cdot\vect r_i}.
\eeq
For simplicity, let us consider a system without two-body interactions, such that the 111-expansion for the
wave function at zero collision energy and zero magnetic quantum number, \Eq{eq:111expansion}, is simplified as
\beq\label{eq:111expansion_no_twobody}
\Psi\approx s_z\left(1 -\frac{3\sqrt{3}D}{8\pi^3B^6}\right).
\eeq
$\widetilde{\Psi}^{(E)}_{\vect{q}_1\vect{q}_2\vect{q}_3}$ can be expressed as \cite{Tan2008}
\begin{widetext}\begin{align}\label{eq:tmatrixdef}
    \widetilde{\Psi}^{(E)}_{\vect{q}_1\vect{q}_2\vect{q}_3} =&\, \frac{(2\pi)^6}{2}\autobra{\delta(\vect q_1-\vect k_1)\delta(\vect q_2-\vect k_2)-\delta(\vect q_1-\vect k_2)\delta(\vect q_2-\vect k_1)}+\frac{1}{2}G^E_{q_1q_2q_3}T(\vect{k}_1\vect{k}_2\vect{k}_3; \vect{q}_1\vect{q}_2\vect{q}_3) \nonumber \\
    & + \autoparen{\text{terms~that~are~regular~at~}\frac{q_1^2+q_2^2+q_3^2}{2}=\frac{m_FE}{\hbar^2}},
\end{align}\end{widetext}
where $T(\vect{k}_1\vect{k}_2\vect{k}_3; \vect{q}_1\vect{q}_2\vect{q}_3)$ is the three-body $T$-matrix element, and $G^E_{q_1q_2q_3}=[(q_1^2+q_2^2+q_3^2)/2-m_FE/\hbar^2-\I\epsilon]^{-1}$, in which the term $-\I\epsilon$ with $\epsilon\to0^+$ specifies an outgoing wave.
The Fourier transform of the incoming wave function $\widetilde{\Psi}_{\text{in}}$ (the first term on the right hand side of \Eq{eq:tmatrixdef}) is $\Psi_{\text{in}} = (e^{\I\vect k_1\cdot\vect r_1}e^{\I\vect k_2\cdot\vect r_2}-e^{\I\vect k_1\cdot\vect r_2}e^{\I\vect k_2\cdot\vect r_1})e^{\I\vect k_3\cdot\vect r_3}/2$. 
Let $k\equiv\sqrt{2m_F E}/\hbar$.
For low-energy scattering such that $k_1\sim k_2\sim k_3\sim k\ll1/r_e$, the de Broglie wave length of each incoming fermion is of the order $\lambda\sim 1/k\gg r_e$. If the hyperradius $B$ satisfies $B\ll\lambda$, $\Psi_{\text{in}}$ can be approximated as
\begin{align}
    \Psi_{\text{in}} &\simeq \I \frac{\vect{k}_1-\vect k_2}{2}\cdot (\vect r_1-\vect r_2).
\end{align}
Thus, according to \Eq{eq:111expansion_no_twobody}, if $r_e\ll B\ll\lambda$, the real-space three-body wave function
may be approximated as
\begin{align}\label{eq:psiout}
    \Psi^{(E)}(\vect r_1,\vect r_2,\vect r_3)&\approx \I \frac{\vect{k}_1-\vect k_2}{2}\cdot (\vect r_1-\vect r_2)\autoparen{1-\frac{3\sqrt{3}D}{8\pi^3B^6}}.
\end{align}
The inverse Fourier transform of the above result is
\begin{align}
    \Psi^{(E)}_{\vect q_1\vect q_2\vect q_3} \approx& \frac{(2\pi)^6}{2}(\vect{k}_1-\vect k_2)\cdot(\nabla_{\vect q_2}-\nabla_{\vect q_1})\delta(\vect q_1)\delta(\vect q_2) \nonumber \\
    &-\frac{D}{2}(\vect{k}_1-\vect k_2)\cdot\frac{\vect{q}_1-\vect q_2}{q_1^2+q_2^2+q_3^2},
\end{align}
which should be consistent with \Eq{eq:tmatrixdef} to the first order in $k$ at $k\to0$. So we have
\begin{align}
    T(\vect{k}_1\vect{k}_2\vect{k}_3; \vect{q}_1\vect{q}_2\vect{q}_3) &\approx -\frac{D}{2}(\vect{k}_1-\vect k_2)\cdot(\vect{q}_1-\vect q_2)
\end{align}
for low-energy three-body collision processes, if there are \emph{no} two-body interactions.
If there are two-body interactions, we conjecture that the $T$-matrix element of three two-component fermions depends linearly on $D$ at low energy, in a way analogous to what was found in Ref.~\cite{Tan2008}.
The detailed computation of $T(\vect{k}_1\vect{k}_2\vect{k}_3; \vect{q}_1\vect{q}_2\vect{q}_3)$ in the presence
of two-body interactions is a future research subject.

\section{\label{sec:born_approximation}The Born Approximation}
If the interaction potentials are weak, we can expand $D$ in powers of the potentials. We first express the three-body wave function as a Born series for weak potentials:
\begin{equation}
    \Psi = \Psi_0 + \widehat{G}\mathcal{V}\Psi_0 + \autoparen{\widehat{G}\mathcal{V}}^2\Psi_0 + \dots,
\end{equation}
where $\mathcal{V}=U(s_1,s_2,s_3)+\sum_iV_i(s_i)$ is the total interaction operator, $\widehat{G}=-\widehat{H}_0^{-1}$ is the Green's operator, and $\widehat{H}_0$ is the three-body kinetic energy operator. 
Starting from \Eq{Psi0}, we derive
\begin{align}
    \widehat{G}\mathcal{V}\Psi_0 =& -\frac{1}{3}s_z\frac{\alpha_{3,3}}{s^3} +\frac{1}{6}s_{1z}\frac{\alpha_{3,1}}{s_1^3} +\frac{1}{6}s_{2z}\frac{\alpha_{3,2}}{s_2^3} - R_{1z}\frac{\alpha_{1,1}}{s_1} \nonumber \\
    & + R_{2z}\frac{\alpha_{1,2}}{s_2} -\frac{\sqrt{3}}{\pi}\frac{s_z}{B^6}\Lambda + O(B^{-6})
\end{align}
at large $B$, where 
\begin{align}
    \alpha_{n,i}\equiv&\frac{m_F}{\hbar^2}\int_0^\infty s'^{n+1}V_i(s')\diff s',\\
    \Lambda\equiv&\, \frac{m_F}{\hbar^2}\int_0^\infty\diff s'\,s'^4\int_0^\infty\diff R'\,R'^2\int_0^{\pi}\diff\theta'\sin\theta' U(\vect s', \vect R'),
\end{align}
where $\theta'$ is the angle between $\vect s'$ and $\vect R'$, and $U(\vect s,\vect R)\equiv U(s_1,s_2,s_3)$.
Comparing these results with the 111-expansion, we find the expansions of $a_0$, $a_1$, $\Tilde{a}_1$ and $D$ to leading order in powers of the potential:
\begin{subequations}\label{eq:bornscatteringpara}
\begin{align}
    a_0 =&\,\alpha_{1,1}=\alpha_{1,2}, \\
    a_1 =&\, \frac{1}{9}\alpha_{3,1}=\frac{1}{9}\alpha_{3,2}, \\
    \Tilde{a}_1 =&\, \frac{1}{9}\alpha_{3,3}, \\
    D =&\, \frac{8\pi^2}{3}\Lambda. \label{eq:bornthreebodypara}
\end{align}
\end{subequations}
The details of the derivation can be found in Appendix \ref{app:born_approximation}.

For any particular two-body potentials $V_i(s_i)$, such as the square-well potential and the Gaussian potential, one can calculate $a_0$, $a_1$ and $\Tilde{a}_1$ by solving the two-body Schr\"{o}dinger equation and verify that the results are consistent with Eqs. \eqref{eq:bornscatteringpara} if the potentials are weak. From \Eq{eq:bornthreebodypara} we know that $D$ linearly depends on $U$ if $U$ is weak. One can find more precise approximate formulas for the two-body scattering parameters and $D$ if the wave function is expanded to the second order in the Born series. When the interactions are not weak, one can solve the three-body Schr\"{o}dinger equation numerically and match the resultant wave function with the asymptotic expansions in \Eq{eq:111expansion} or \Eq{eq:2nd21expansion} to extract the numerical value of $D$.

\section{\label{sec:energy_shift}Shifts of energy due to $D$}
In this section we first consider one spin-$\uparrow$ fermion and two spin-$\downarrow$ fermions in a periodic box and study the energy shifts of this system caused by the three-body scattering hypervolume. We then derive the thermodynamic properties of two-component Fermi gases due to the nonzero three-body scattering hypervolume.

\subsection{One spin-$\uparrow$ fermion and two spin-$\downarrow$ fermions in a cubic box}
For simplicity, we assume that all the two-body interactions are fine-tuned such that the two-body scattering
phase shifts are zero at small collision energies, but the three-body hypervolume $D\ne0$. The particles are labeled in
the same way as in Sec. \ref{sec:asymptotics}.
We place them in a large periodic cubic box with volume $\Omega$.
If these fermions have no interactions, we may consider an energy eigenstate in which fermions 1, 2, and 3 have momenta
 $\hbar\mathbf{k}_1$, $\hbar\mathbf{k}_2$, and $\hbar\mathbf{k}_3$ respectively, with normalized wave function
\begin{equation}
    \Psi_{\mathbf{k}_1,\mathbf{k}_2,\mathbf{k}_3}(\vect r_1,\vect r_2,\vect r_3) = \frac{e^{\I\mathbf{k}_3\cdot \mathbf{r}_3}}{\sqrt{2}\Omega^{3/2}}\begin{vmatrix}
        e^{\I\mathbf{k}_1\cdot \mathbf{r}_1} & e^{\I\mathbf{k}_1\cdot \mathbf{r}_2}\\
        e^{\I\mathbf{k}_2\cdot \mathbf{r}_1} & e^{\I\mathbf{k}_2\cdot \mathbf{r}_2}
    \end{vmatrix}.
\end{equation}
We define the momenta $\hbar\mathbf{k}_c$, $\hbar\mathbf{p}$ and $\hbar\mathbf{q}$ in the COM frame such that
\begin{align}
    \mathbf{k}_1 =&\, \frac{1}{3}\mathbf{k}_c + \frac{1}{2}\mathbf{q} + \mathbf{p}, \\
    \mathbf{k}_2 =&\, \frac{1}{3}\mathbf{k}_c + \frac{1}{2}\mathbf{q} - \mathbf{p}, \\
    \mathbf{k}_3 =&\, \frac{1}{3}\mathbf{k}_c - \mathbf{q},
\end{align}
where $\hbar\mathbf{k}_c$ is the total momentum.
We then factorize the wave function as 
\begin{equation}
    \Psi_{\mathbf{k}_1,\mathbf{k}_2,\mathbf{k}_3}(\vect r_1,\vect r_2,\vect r_3)  = \frac{1}{\sqrt{\Omega}}e^{\I\mathbf{k}_c\cdot\mathbf{R}_c}\Phi_{\mathbf{q}, \mathbf{p}},
\end{equation}
where $\mathbf{R}_c=(\mathbf{r}_1+\mathbf{r}_2+\mathbf{r}_3)/3$, and 
\begin{equation}
    \Phi_{\mathbf{q}, \mathbf{p}} = \frac{1}{\sqrt{2}\Omega}e^{-\I\mathbf{q}\cdot\mathbf{R}_3}\autoparen{e^{\I\mathbf{p}\cdot\mathbf{s}_3}-e^{-\I\mathbf{p}\cdot\mathbf{s}_3}}
\end{equation}
is the wave function for the relative motion of the three particles. 
Consider a state with $k_1\sim k_2\sim k_3\sim k\ll1/r_e$, in which the de Broglie wave length of each
fermion is of the order of $\lambda=2\pi/k\gg r_e$.
When we introduce short-range interactions with range $r_e$, the wave function is only slightly modified at $B\gg r_e$.
When the hyperradius $B$ satisfies $r_e\ll B\ll\lambda$, one can expand $\Phi_{\mathbf{q}, \mathbf{p}}$ as
\begin{equation}
    \Phi_{\mathbf{q}, \mathbf{p}} \simeq \frac{\sqrt{2}}{\Omega}\I\mathbf{p}\cdot\mathbf{s}_3.
\end{equation}
We then introduce a nonzero $D$ adiabatically, and $\Phi_{\mathbf{q}, \mathbf{p}}$ is changed to
\begin{equation}
    \Phi_{\mathbf{q}, \mathbf{p}} \simeq \frac{\sqrt{2}}{\Omega}\I\mathbf{p}\cdot\mathbf{s}_3\autoparen{1-\frac{3\sqrt{3}D}{8\pi^3B^6}}
\end{equation}
for $r_e\ll B\ll\lambda$.

$\Phi_{\mathbf{q}, \mathbf{p}}$ satisfies the Schr\"{o}dinger equation,
\begin{equation}
    -\frac{\hbar^2}{m_F}\nabla^2_{\brho} \Phi_{\mathbf{q}, \mathbf{p}} = E\Phi_{\mathbf{q}, \mathbf{p}},
\end{equation}
outside the range of interaction, where $E$ is the energy of the relative motion, and $\brho=(\mathbf{s}, 2\mathbf{R}/\sqrt{3})$. Suppose that the box size $\Omega^{1/3}$ is large, and consider two different interactions that yield two different three-body scattering hypervolumes $D_1$ and $D_2$. We get the following two equations
\begin{subequations}
\begin{align}
    -\frac{\hbar^2}{m_F}\nabla^2_{\brho} \Phi_1 =&\, E_1\Phi_1, \label{eq:phiSchrodinger1} \\
    -\frac{\hbar^2}{m_F}\nabla^2_{\brho} \Phi_2 =&\, E_2\Phi_2, \label{eq:phiSchrodinger2}
\end{align}
\end{subequations}
where the subscripts $\mathbf{q}, \mathbf{p}$ have been suppressed for simplicity.
Multiplying both sides of \Eq{eq:phiSchrodinger1} by $\Phi_2^*$, multiplying both sides of \Eq{eq:phiSchrodinger2} by $\Phi_1^*$ and taking the complex conjugates of both sides, subtracting the two resultant equations, and taking the integral over $\brho$ for $\rho>\rho_0$, we find
\begin{align}\label{eq:phiint}
    \autoparen{E_1-E_2} &\int_{\rho>\rho_0}\diff^6\rho\,\Phi_1\Phi_2^* \nonumber \\
    & =-\frac{\hbar^2}{m_F}\int_{\rho>\rho_0}\diff^6\rho\,\autoparen{\Phi^*_2\nabla^2_{\brho}\Phi_1-\Phi_1\nabla^2_{\brho}\Phi^*_2},
\end{align}
where $\rho_0$ is a length scale such that $r_e\ll\rho_0\ll\lambda$. Note that $\Phi_1\simeq\Phi_2$ in the region $\rho>\rho_0$, and the volume of the region $\rho<\rho_0$ is small and can be neglected on the left hand side of \Eq{eq:phiint}, so the left-hand side of \Eq{eq:phiint} becomes
\begin{equation}
    \frac{8}{3\sqrt{3}}\autoparen{E_1-E_2}\int_{\rho>\rho_0}\diff^3s\,\diff^3R\, |\Phi|^2 \simeq \frac{8}{3\sqrt{3}}\autoparen{E_1-E_2}.
\end{equation}
We then carry out the integral on the right-hand side of \Eq{eq:phiint} by applying Gauss's divergence theorem and get
\begin{equation}
    E_2-E_1 = \frac{\hbar^2\autoparen{D_2-D_1}}{2m_F\Omega^2}\autoparen{\mathbf{k}_1-\mathbf{k}_2}^2.
\end{equation}
This result implies that the energy shift $\mathcal{E}_{\mathbf{k}_1, \mathbf{k}_2, \mathbf{k}_3}$ of such a three-body system due to the three-body scattering hypervolume is
\begin{equation}\label{eq:3bodyenergyshift}
    \mathcal{E}_{\mathbf{k}_1, \mathbf{k}_2, \mathbf{k}_3} = \frac{\hbar^2 D}{2m_F\Omega^2}\autoparen{\mathbf{k}_1-\mathbf{k}_2}^2.
\end{equation}

\subsection{Energy shifts of many two-component fermions and the thermodynamic consequences}
Now one can generalize \Eq{eq:3bodyenergyshift} to $N$ fermions in the large periodic box with volume $\Omega$. Suppose there are $N_\sigma$ spin-$\sigma$ fermions, where $\sigma=\uparrow, \downarrow$, and $N=N_\uparrow+N_\downarrow$. The number density of the spin-$\sigma$ fermions is $n_\sigma=N_\sigma/\Omega$, and we assume the density is low enough such that the average interparticle distance $n^{-1/3}\equiv(n_\uparrow+n_\downarrow)^{-1/3}\gg r_e$. For later use, we define the Fermi wave number $k_{F\sigma}\equiv(6\pi^2n_\sigma)^{1/3}$, the Fermi energy $\epsilon_{F\sigma}\equiv\hbar^2k_{F\sigma}^2/2m_F$, and the Fermi temperature $T_{F\sigma}\equiv\epsilon_{F\sigma}/k_B$, where $k_B$ is the Boltzmann constant. 

\subsubsection{Adiabatic shifts of energy in the thermodynamic limit}
Note that in such a system, there are \textit{four} different scattering hypervolumes, namely, $D_{\uparrow\uparrow\uparrow}$, $D_{\uparrow\uparrow\downarrow}$, $D_{\downarrow\downarrow\uparrow}$, and $D_{\downarrow\downarrow\downarrow}$, where the subscript $\sigma\sigma'\sigma''$ means that the three-body scattering hypervolume is defined to describe the scattering between three particles with spins $\sigma$, $\sigma'$, and $\sigma''$. $D_{\uparrow\uparrow\uparrow}$ and $D_{\downarrow\downarrow\downarrow}$ have been studied in Ref.~\cite{Wang2021-2}. If we introduce these four nonzero scattering hypervolumes adiabatically, the energy shift at the first order in the scattering hypervolumes is equal to the sum of the contributions from all the triples of fermions with spins $\uparrow\uparrow\uparrow$, $\uparrow\uparrow\downarrow$, $\downarrow\downarrow\uparrow$, or $\downarrow\downarrow\downarrow$. We find
\begin{align}
    \Delta E =&\, \sum_{\mathbf{k}_1\mathbf{k}_2\mathbf{k}_3} \left(\frac{1}{2!}\mathcal{E}_{\mathbf{k}_1\mathbf{k}_2\mathbf{k}_3}^{D_{\uparrow\uparrow\downarrow}} n_{\mathbf{k}_1\downarrow}n_{\mathbf{k}_2\uparrow}n_{\mathbf{k}_3\uparrow}\right. \nonumber \\
    &\, + \frac{1}{2!}\mathcal{E}_{\mathbf{k}_1\mathbf{k}_2\mathbf{k}_3}^{D_{\downarrow\downarrow\uparrow}} n_{\mathbf{k}_1\uparrow}n_{\mathbf{k}_2\downarrow}n_{\mathbf{k}_3\downarrow} \nonumber \\
    &\, + \frac{1}{3!}\mathcal{E}_{\mathbf{k}_1\mathbf{k}_2\mathbf{k}_3}^{D_{\uparrow\uparrow\uparrow}} n_{\mathbf{k}_1\uparrow}n_{\mathbf{k}_2\uparrow}n_{\mathbf{k}_3\uparrow} \nonumber \\
    &\, \left. +\, \frac{1}{3!}\mathcal{E}_{\mathbf{k}_1\mathbf{k}_2\mathbf{k}_3}^{D_{\downarrow\downarrow\downarrow}} n_{\mathbf{k}_1\downarrow}n_{\mathbf{k}_2\downarrow}n_{\mathbf{k}_3\downarrow}\right),
\end{align}
where $n_{\mathbf{k}\sigma}=\autoparen{1+e^{\beta\autoparen{\epsilon_{\mathbf{k}}-\mu_\sigma}}}^{-1}$ is the Fermi-Dirac distribution function for spin-$\sigma$ fermions, $\beta=1/k_BT$, $T$ is the temperature of the system, $\epsilon_{\mathbf{k}}=\hbar^2k^2/2m_F$ is the kinetic energy of a spin-$\sigma$ fermion with momentum $\hbar\mathbf{k}$, and $\mu_\sigma$ is the chemical potential of spin-$\sigma$ fermions.
The energy shifts due to $D_{\uparrow\uparrow\uparrow}$ and $D_{\downarrow\downarrow\downarrow}$ have been studied in Ref.~\cite{Wang2021-2}.
In the thermodynamic limit, the summation over $\mathbf{k}$ can be replaced by the integral over $\mathbf{k}$, i.e., $\sum_\mathbf{k}\rightarrow\Omega\int\diff^3k/(2\pi)^3$. We carry out the integral and find
\begin{align}
    \Delta E =&\, -\frac{\sqrt{\pi}\hbar^2}{64\pi^{4}m_F}\sum_\sigma N_{\bar{\sigma}} D_{\sigma\sigma\bar{\sigma}} k^8_{F\sigma}\widetilde{T}^{5/2}_{\sigma}\mathrm{Li}_{5/2}\autoparen{-e^{\widetilde{\mu}_\sigma/\widetilde{T}_{\sigma}}} \nonumber \\
    &\, +\frac{\hbar^2}{256\pi^{3}m_F}\sum_\sigma N_{\sigma} D_{\sigma\sigma\sigma} k^{10}_{F\sigma}\widetilde{T}^{5}_{\sigma}\mathrm{Li}^2_{5/2}\autoparen{-e^{\widetilde{\mu}_\sigma/\widetilde{T}_{\sigma}}},
\end{align}
where $\bar{\sigma}=\,\downarrow$ when $\sigma=\,\uparrow$, and $\bar{\sigma}=\,\uparrow$ when $\sigma=\,\downarrow$, $\widetilde{T}_{\sigma}=T/T_{F\sigma}$, $\widetilde{\mu}_\sigma=\mu_\sigma/\epsilon_{F\sigma}$, and $\mathrm{Li}_{\nu}(z)$ is the polylogarithm function of order $\nu$. The chemical potential $\mu_\sigma$ can be determined by the number of fermions, i.e.,
\begin{equation}
    N_\sigma = \Omega\int\frac{\diff^3k}{(2\pi)^3}\frac{1}{1+e^{\beta\autoparen{\epsilon_{\mathbf{k}}-\mu_\sigma}}},
\end{equation}
which leads to the following equation:
\begin{equation}
    1=-\frac{3\sqrt{\pi}}{4}\widetilde{T}^{3/2}_{\sigma}\mathrm{Li}_{3/2}\autoparen{-e^{\widetilde{\mu}_\sigma/\widetilde{T}_{\sigma}}}.
\end{equation}

In the low temperature limit, $T\ll T_{F\sigma}$, we get
\begin{align}
    \Delta E(T) =& \sum_\sigma\left[\frac{\hbar^2N_{\bar{\sigma}}D_{\sigma\sigma\bar{\sigma}} k^8_{F\sigma}}{24\pi^4m_F}  \autoparen{\frac{1}{5}+\frac{\pi^2}{12}\widetilde{T}^{2}_{\sigma}+O(\widetilde{T}^{4}_{\sigma})} \right.\nonumber \\
    & \left.+ \frac{\hbar^2N_{\sigma}D_{\sigma\sigma\sigma}k^{10}_{F\sigma}}{180\pi^4m_F} \autoparen{\frac{1}{5}+\frac{\pi^2}{6}\widetilde{T}^{2}_{\sigma}+O(\widetilde{T}^{4}_{\sigma})}\right].
\end{align}
In particular, at zero temperature we get
\begin{align}
    \Delta E(0) =&\, \frac{\hbar^2}{120\pi^4m_F}\sum_\sigma N_{\bar{\sigma}} D_{\sigma\sigma\bar{\sigma}} k^8_{F\sigma} \nonumber \\
    &\, + \frac{\hbar^2}{900\pi^4m_F}\sum_\sigma N_{\sigma} D_{\sigma\sigma\sigma} k^{10}_{F\sigma}.
\end{align}
If we consider the effects of $D_{\sigma\sigma\bar{\sigma}}$ only, the energy shift is
\begin{equation}
    \Delta E(T) = \frac{\hbar^2}{24\pi^4m_F}\sum_\sigma N_{\bar{\sigma}} D_{\sigma\sigma\bar{\sigma}} k^8_{F\sigma}\autoparen{\frac{1}{5}+\frac{\pi^2}{12}\widetilde{T}^{2}_{\sigma}+O(\widetilde{T}^{4}_{\sigma})}
\end{equation}
at $T\ll T_{F\sigma}$.

In the intermediate temperature regime, $T_{F\sigma}\ll T\ll T_e$, we have
\begin{align}
    \Delta E(T) =&\, \frac{\hbar^2}{48\pi^4m_F}\sum_\sigma N_{\bar{\sigma}} D_{\sigma\sigma\bar{\sigma}} k^8_{F\sigma} \nonumber \\
    &\,\times\autoparen{\widetilde{T}_\sigma+\frac{1}{3\sqrt{2\pi}}\widetilde{T}^{-1/2}_{\sigma}+O(\widetilde{T}^{-2}_{\sigma})} \nonumber \\
    &\, + \frac{\hbar^2}{72\pi^4m_F}\sum_\sigma N_{\sigma} D_{\sigma\sigma\sigma} k^{10}_{F\sigma} \nonumber \\
    &\, \times\autoparen{\frac{1}{2}\widetilde{T}^2_\sigma+\frac{1}{3\sqrt{2\pi}}\widetilde{T}^{1/2}_{\sigma}+O(\widetilde{T}^{-1}_{\sigma})},
\end{align}
where $T_e\equiv\frac{\hbar^2}{2m_Fr^2_ek_B}$. The energy shift due to $D_{\sigma\sigma\bar{\sigma}}$ is
\begin{align}
    \Delta E(T) =&\, \frac{\hbar^2}{48\pi^4m_F}\sum_\sigma N_{\bar{\sigma}} D_{\sigma\sigma\bar{\sigma}} k^8_{F\sigma} \nonumber \\
    &\,\times\autoparen{\widetilde{T}_\sigma+\frac{1}{3\sqrt{2\pi}}\widetilde{T}^{-1/2}_{\sigma}+O(\widetilde{T}^{-2}_{\sigma})}
\end{align}
at $T_{F\sigma}\ll T\ll T_e$.

\subsubsection{Isothermal shifts of energy in the thermodynamic limit}
For the sake of simplicity, we now assume that only $D_{\downarrow\downarrow\uparrow}$ is nonzero, but $D_{\uparrow\uparrow\downarrow}$, $D_{\uparrow\uparrow\uparrow}$, and $D_{\downarrow\downarrow\downarrow}$ are zero. If such an interaction is introduced adiabatically, the change of temperature will be
\begin{equation}
    \Delta T = \autoparen{\frac{\partial\Delta E}{\partial S}}_{N,\Omega},
\end{equation}
and the temperature will increase (if $D_{\downarrow\downarrow\uparrow}>0$) or decrease (if $D_{\downarrow\downarrow\uparrow}<0$). 
Thus if $D_{\downarrow\downarrow\uparrow}$ is introduced isothermally, the shift of energy $\Delta E'$ should be approximately
\begin{equation}
    \Delta E' = \Delta E - C\Delta T = \autoparen{1-T\frac{\partial}{\partial T}}\Delta E,
\end{equation}
where $C$ is the heat capacity of the non-interacting two-component Fermi gases at constant volume. In the low temperature limit, $T\ll T_{F\sigma}$,
\begin{equation}
    \Delta E'(T) = \frac{\hbar^2N_\uparrow D_{\downarrow\downarrow\uparrow}}{24\pi^4m_F} k^8_{F\downarrow}\autoparen{\frac{1}{5}-\frac{\pi^2}{12}\widetilde{T}^2_{\downarrow}+O(\widetilde{T}^4_{\downarrow})}.
\end{equation}
In the intermediate temperature regime, $T_{F\sigma}\ll T\ll T_e$, 
\begin{equation}
    \Delta E'(T) = \frac{\hbar^2N_\uparrow D_{\downarrow\downarrow\uparrow}}{48\pi^4m_F} k^8_{F\downarrow}\autoparen{\frac{1}{2\sqrt{2\pi}}\widetilde{T}_\downarrow^{-1/2}+O(\widetilde{T}_\downarrow^{-2})}.
\end{equation}

\begin{figure}[t]
    \centering
    \includegraphics[width=\linewidth]{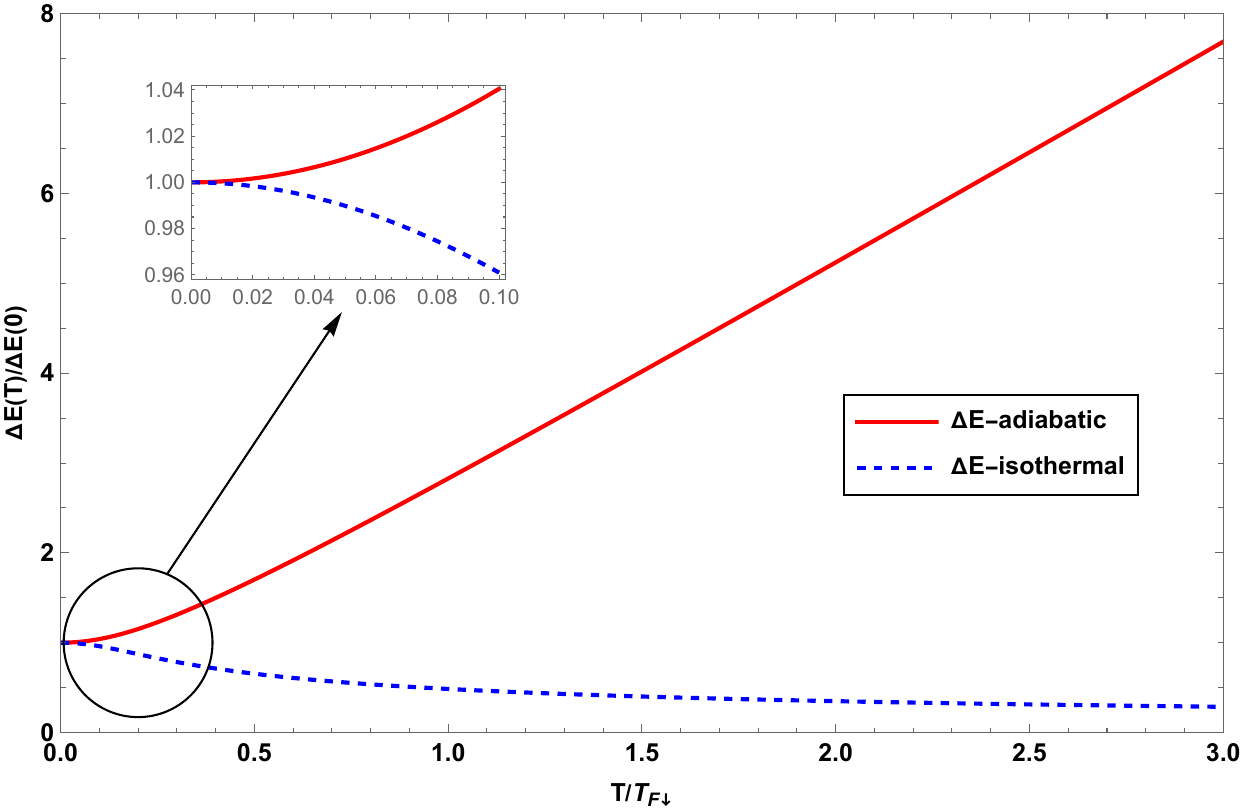}
    \caption{The shifts of energy caused by the adiabatic (red line) or isothermal (blue dashed line) introduction of $D_{\downarrow\downarrow\uparrow}$ as functions of temperature.}
    \label{fig:energyshifts}
\end{figure}

We plot the shifts of energy as functions of temperature in Fig.~\ref{fig:energyshifts}.

\section{\label{sec:three-body_recomb}Inelastic scatterings and three-body recombination rate}
In Sec. \ref{sec:asymptotics}, we mentioned that the three-body scattering hypervolume is real if the scattering is purely elastic, but it will gain a negative imaginary part if the scattering is inelastic. In the latter case, the interactions can support one or more two-body bound states, and the three-body recombination will occur. For most ultracold atomic gases, the three-body recombination is possible, since most ultracold atoms have two-body bound states, and it will cause atom loss of the ultracold gas. 

When the scattering hypervolume becomes complex, the energy shift also gains a negative imaginary part. Within a short time $\Delta t$, the probability that no three-body recombination occurs is $\exp(-2|\Im E|\Delta t/\hbar)\simeq 1-2|\Im(E)|\Delta t/\hbar$, and the probability of one recombination is $2|\Im(E)|\Delta t/\hbar$. After one recombination event, three atoms will escape from the trap, thus the change of the number of spin-$\sigma$ atoms in a short time $\diff t$ is
\begin{align}
    \diff N_\sigma =&\, -2\frac{\diff t}{\hbar}\sum_{\mathbf{k}_1\mathbf{k}_2\mathbf{k}_3} \left(\frac{1}{2!}\left|\Im\mathcal{E}_{\mathbf{k}_1\mathbf{k}_2\mathbf{k}_3}^{D_{\sigma\bar{\sigma}\bar{\sigma}}}\right| n_{\mathbf{k}_1\sigma}n_{\mathbf{k}_2\bar{\sigma}}n_{\mathbf{k}_3\bar{\sigma}}\right. \nonumber \\
    &\, +2\frac{1}{2!}\left|\Im\mathcal{E}_{\mathbf{k}_1\mathbf{k}_2\mathbf{k}_3}^{D_{\sigma\sigma\bar{\sigma}}}\right| n_{\mathbf{k}_1\sigma}n_{\mathbf{k}_2\sigma}n_{\mathbf{k}_3\bar{\sigma}} \nonumber \\
    &\, \left. +\,3\frac{1}{3!}\left|\Im\mathcal{E}_{\mathbf{k}_1\mathbf{k}_2\mathbf{k}_3}^{D_{\sigma\sigma\sigma}}\right| n_{\mathbf{k}_1\sigma}n_{\mathbf{k}_2\sigma}n_{\mathbf{k}_3\sigma} \right).
\end{align}
This leads to
\begin{align}
    \frac{\diff n_\sigma}{\diff t} = -L_{3\sigma}^\scriptp{1} n_\sigma n_{\bar{\sigma}}^2 - L_{3\sigma}^\scriptp{2} n_\sigma^2 n_{\bar{\sigma}} - L_{3\sigma}^\scriptp{3} n_\sigma^3,
\end{align}
where $L_{3\sigma}^\scriptp{1}$, $L_{3\sigma}^\scriptp{2}$, $L_{3\sigma}^\scriptp{3}$ are the three-body recombination rate constants and they can be expressed as
\begin{subequations}
\begin{align}
    L_{3\sigma}^\scriptp{1} =& -\frac{9\sqrt{\pi}\hbar}{8m_F}\widetilde{T}^{5/2}_{\bar{\sigma}}\mathrm{Li}_{5/2}\autoparen{-e^{\widetilde{\mu}_{\bar{\sigma}}/\widetilde{T}_{\bar{\sigma}}}} \left|  
    \Im D_{\sigma\bar{\sigma}\bar{\sigma}}\right| k_{F\bar{\sigma}}^2, \\  
    L_{3\sigma}^\scriptp{2} =& -\frac{9\sqrt{\pi}\hbar}{4m_F}\widetilde{T}^{5/2}_{\sigma}\mathrm{Li}_{5/2}\autoparen{-e^{\widetilde{\mu}_{\sigma}/\widetilde{T}_{\sigma}}} \left|\Im D_{\sigma\sigma\bar{\sigma}}\right| k_{F\sigma}^2, \\ 
    L_{3\sigma}^\scriptp{3} =&\, \frac{27\pi\hbar}{32m_F}\widetilde{T}^{5}_{\sigma}\mathrm{Li}^2_{5/2}\autoparen{-e^{\widetilde{\mu}_{\sigma}/\widetilde{T}_{\sigma}}} \left|\Im D_{\sigma\sigma\sigma}\right| k_{F\sigma}^4. \label{eq:L33sigma}
\end{align}
\end{subequations}
$L_{3\sigma}^\scriptp{1}$ depends on the density $n_{\bar{\sigma}}$ and temperature $T$, and $L_{3\sigma}^\scriptp{2}$ and $L_{3\sigma}^\scriptp{3}$ depend on the density $n_\sigma$ and temperature $T$.

In the low temperature limit, $T\ll T_{F\sigma}$, we have
\begin{subequations}
\begin{align}
    L_{3\sigma}^\scriptp{1} \simeq&\, \frac{3}{5}\frac{\hbar}{m_F} \left|  
    \Im D_{\sigma\bar{\sigma}\bar{\sigma}}\right| k_{F\bar{\sigma}}^2 \autoparen{1+\frac{5\pi^2}{12}\widetilde{T}^2_{\bar{\sigma}}}, \\
    L_{3\sigma}^\scriptp{2} \simeq&\, \frac{6}{5}\frac{\hbar}{m_F} \left|  
    \Im D_{\sigma\sigma\bar{\sigma}}\right| k_{F\sigma}^2 \autoparen{1+\frac{5\pi^2}{12}\widetilde{T}^2_{\sigma}}, \\
    L_{3\sigma}^\scriptp{3} \simeq&\, \frac{6}{25}\frac{\hbar}{m_F} \left|  
    \Im D_{\sigma\sigma\sigma}\right| k_{F\sigma}^4 \autoparen{1+\frac{5\pi^2}{6}\widetilde{T}^2_{\sigma}}. \label{eq:L33sigma1}
\end{align}
\end{subequations}
In the intermediate temperature regime, $ T_{F\sigma}\ll T\ll T_e $, we have
\begin{subequations}\label{eq:threerecombapprox}
\begin{align}
    L_{3\sigma}^\scriptp{1} \simeq&\, \frac{3}{\hbar}\left|  
    \Im D_{\sigma\bar{\sigma}\bar{\sigma}}\right| \autoparen{k_B T}, \\
    L_{3\sigma}^\scriptp{2} \simeq&\, \frac{6}{\hbar} \left|
    \Im D_{\sigma\sigma\bar{\sigma}}\right| \autoparen{k_B T}, \\
    L_{3\sigma}^\scriptp{3} \simeq&\, 6\frac{m_F}{\hbar^3} \left|  
    \Im D_{\sigma\sigma\sigma}\right|  \autoparen{k_B T}^2. \label{eq:L33sigma2}
\end{align}
\end{subequations}
Equations~(\ref{eq:L33sigma}), (\ref{eq:L33sigma1}), and (\ref{eq:L33sigma2}) were first derived in Ref.~\cite{Wang2021-2}.

According to the three-body threshold laws, the three-body recombination rate constants $L_{3\sigma}^\scriptp{1}$ and $L_{3\sigma}^\scriptp{2}$ are predicted to be proportional to $T^1$ \cite{Esry2001, Esry2005}, and $L_{3\sigma}^\scriptp{3}$ is predicted to be proportional to $T^2$ \cite{Esry2001, Esry2005}. Experimentally, these predictions are confirmed in $^6$Li gases \cite{Yoshida2019, Ketterle2021, Ji2022}. Our results Eqs.~\eqref{eq:threerecombapprox} are consistent with these results.

\section{\label{sec:summary}Summary and discussion}
We studied the three-body problem for two-component fermions (one spin-$\uparrow$ fermion and two spin-$\downarrow$ fermions)
in 3D and derived three asymptotic expansions for the wave function $\Psi$. The scattering energy is assumed to be zero and the angular momentum quantum number $L$ is assumed to be 1. We defined a new three-body scattering hypervolume $D$, which appears at the order of $B^{-5}$ in the 111-expansion and has the dimension of length to the sixth power. The three-body scattering hypervolume plays an important role in the low energy physics of three or more fermions.

We studied the $T$-matrix element for three fermions (one spin-$\uparrow$ fermion and two spin-$\downarrow$ fermions) colliding at a small nonzero energy. In the absence of two-body interactions we found a simple formula for this $T$-matrix element in terms of $D$.

We also derived an asymptotic formula for $D$ by using the Born expansion when the interactions are weak. If the interactions are strong, one may numerically solve the three-body Schr\"{o}dinger equation and extract the numerical value of $D$ by matching the numerical wave function and the asymptotic expansions we have derived.

In the remaining part of this paper, we considered these three two-component fermions in a large box with periodic boundary condition imposed on the wave function, and found the shifts of the energy eigenvalues due to a nonzero $D$. We then studied the dilute two-component Fermi gases and derived the shifts of their energies due to the four three-body scattering hypervolumes ($D_{\uparrow\uparrow\uparrow}$, $D_{\uparrow\uparrow\downarrow}$, $D_{\downarrow\downarrow\uparrow}$, and $D_{\downarrow\downarrow\downarrow}$). 

If the interactions can support one or more two-body bound states, then the three-body recombination will occur and $D$ must gain a negative imaginary part to satisfy the conservation of probability. We derived formulas for the three-body recombination rate constants $L_{3\sigma}^\scriptp{1}$, $L_{3\sigma}^\scriptp{2}$ and $L_{3\sigma}^\scriptp{3}$ in terms of the imaginary parts of the three-body scattering hypervolumes, the temperature and the densities of the two components of the Fermi gas.

When there is a bound state of three such fermions with energy close to zero, there will be three-body resonance for low energy three-body collisions. If there is a three-body bound state whose orbital angular momentum quantum number is 1 and whose energy is close to zero, we expect that the the three-body scattering hypervolume $D$ defined in this paper will be anomalously large,
causing strong effects in few-body and many-body physics. If the energy of such a three-body bound state is negative and close to zero, we expect that $\Re D$ is large and positive; but if this three-body bound state is a metastable one with \emph{positive} and small energy, we expect that $\Re D$ is large in magnitude and negative; in both cases, $|\Im D|$ should be large as well, if there are deeper two-body bound states such that the system can undergo three-body recombination.

One can generalize the three-body scattering hypervolume defined in this paper to any three-body systems containing two identical fermions in the same spin state and one different particle. If the different particle has the same mass as each of the identical fermions, the three-body scattering hypervolume defined in this paper and the 111- and the 21-expansions of the zero-energy wave function derived in this paper will remain applicable. But if the different particle has a different mass compared to each of the two identical fermions, one will have to rederive the 111- and the 21-expansions, but one will still have the three-body scattering hypervolume in these expansions, and its dimension should still be length to the sixth power.


In many ultracold atomic simulations of quantum many-body models such as the Fermi Hubbard model, ultracold fermionic atoms are used. We expect that the three-body scattering hypervolume considered in this paper will cause at least tiny effects in the experimental results.

The three-body scattering hypervolume studied in this paper is applicable to not only ultracold atoms, but also to other particles
with short-range interactions such as neutrons. Although people have measured the two-body scattering length and effective range and shape parameters of neutrons, to our knowledge nobody has measured the three-body scattering hypervolumes of the neutrons,
nor has anybody measured the three-body scattering hypervolumes of two neutrons and a third particle (such as a proton, another atomic nucleus, an electron, an atom, or an ion). In future precision studies of material science using high-density neutron beams as probes, the three-body scattering hypervolumes might become important.

\appendix
\section{\label{app:twobodyfunctions}TWO-BODY SPECIAL FUNCTIONS}
Here we introduce the two-body special functions \cite{Tan2008} $\phi^\script{l}{m}(\vect s_1)$, $f^\script{l}{m}(\vect s_1)$, $g^\script{l}{m}(\vect s_1)$, $\dots$, and $\tphi^\script{l}{m}(\vect s)$, $\tf^\script{l}{m}(\vect s)$, $\tg^\script{l}{m}(\vect s)$, $\dots$ that will be used in the 21-expansions.

Consider the scattering between one spin-$\uparrow$ fermion and one spin-$\downarrow$ fermion in the COM reference frame with collision energy $E=\hbar^2k^2/(2\mu)$,
orbital angular momentum quantum number $l$ and magnetic quantum number $m$ along the $z$ direction,
where $\mu=m_F/2$ is the reduced mass.
The wave function $\psi^\script{l}{m}(\vect s_1)$ can be separated as 
\begin{equation}\label{eq:psiseparate}
    \psi^\script{l}{m}(\vect s_1)=u(s_1)\sqrt{\frac{4\pi}{2l+1}}Y_l^m(\hat{\vect{s}}_1),
\end{equation}
where $Y_l^m(\hat{\vect{s}}_1)$ is the spherical harmonics \cite{SH}.
$u(s_1)$ satisfies the radial Schr\"{o}dinger equation
\begin{equation}\label{eq:radialeq}
    \frac{\diff^2u}{\diff s_1^2}+\frac{2}{s_1}\frac{\diff u}{\diff s_1}+\left[k^2-\frac{2\mu V_1(s_1)}{\hbar^2}-\frac{l(l+1)}{s_1^2}\right]u=0.
\end{equation}
Since the interaction potential $V_1(s_1)$ is finite-ranged, we can find the analytical solution for $u(s_1)$ outside of the range of interaction:
\begin{equation}\label{eq:radialwave}
    u(s_1)=\alpha(l, k)\left[j_l(ks_1)\cot\delta_l(k)-y_l(ks_1)\right],~s_1>r_e,
\end{equation}
where $\alpha(l, k)$ is an arbitrary coefficient that determines the overall amplitude of the wave function, $j_l$ and $y_l$ are the spherical Bessel functions of the first and the second kind respectively, and $\delta_l(k)$ is the $l$-wave scattering phase shift
between a spin-$\uparrow$ fermion and a spin-$\downarrow$ fermion. 

In the COM frame, the Schr\"{o}dinger equation for the collision of particles 2 and 3 with
energy $\hbar^2k^2/(2\mu)$ and specified orbital angular momentum quantum numbers may be written as
\begin{equation}
\widetilde{H}_1\psi^{(l,m)}(\vect s_1)=k^2\psi^{(l,m)}(\vect s_1),
\end{equation}
where 
\begin{equation}
    \widetilde{H}_1 \equiv -\nabla^2_{\vect s_1} + \frac{2\mu}{\hbar^2}V_1(s_1)
\end{equation}
is proportional to the two-body Hamiltonian in the COM frame.
For low energy scatterings, we can expand the wave function in powers of $k^2$ \cite{Tan2008}:
\begin{equation}\label{eq:wavefuncexpansion}
    \psi^\script{l}{m}(\vect s_1)=\phi^\script{l}{m}(\vect s_1)+k^2f^\script{l}{m}(\vect s_1)+k^4g^\script{l}{m}(\vect s_1)+\dots,
\end{equation}
and the two-body special functions $\phi^{(l,m)}(\vect s_1)$, $f^{(l,m)}(\vect s_1)$, and $g^{(l,m)}(\vect s_1)$ 
are independent of $k$ and satisfy \cite{Tan2008}
\begin{subequations}\label{eq:Hphi}
\begin{align}
    \widetilde{H}_1\phi^\script{l}{m}(\vect s_1) &=\, 0, \\
    \widetilde{H}_1f^\script{l}{m}(\vect s_1) &=\, \phi^\script{l}{m}(\vect s_1), \\
    \widetilde{H}_1g^\script{l}{m}(\vect s_1) &=\, f^\script{l}{m}(\vect s_1), \dots.
\end{align}
\end{subequations}
 Solving Eqs.~\eqref{eq:Hphi}, we find
that if $s_1>r_e$ we have
\begin{subequations}\label{eq:twobodyspecialfunction}
\begin{align}
    \phi^\script{l}{m}(\vect s_1) =& \autobra{\frac{s_1^l}{(2l+1)!!}-\frac{(2l-1)!!a_l}{s_1^{l+1}}}\sqrt{\frac{4\pi}{2l+1}}\shp{l}{m}{s}, \\
    f^\script{l}{m}(\vect s_1) =& -\autobra{\frac{s_1^{l+2}}{2(2l+3)!!}+\frac{(2l-3)!!a_l}{2s_1^{l-1}}+\frac{a_lr_ls_1^l}{2(2l+1)!!}}\nonumber \\
    & \times\sqrt{\frac{4\pi}{2l+1}}\shp{l}{m}{s}, \\
    g^\script{l}{m}(\vect s_1)=& \left[\frac{s_1^{l+4}}{8(2l+5)!!}-\frac{(2l-5)!!a_l}{8s_1^{l-3}}+\frac{a_lr_ls_1^{l+2}}{4(2l+3)!!}\right.\nonumber \\
    & \left.-\frac{a_lr'_ls_1^l}{24(2l+1)!!}\right]\sqrt{\frac{4\pi}{2l+1}}\shp{l}{m}{s},
\end{align}
\end{subequations}
where we have chosen a particular overall amplitude for $\phi^{(l,m)}(\vect s_1)$ in accordance with Ref.~\cite{Tan2008},
and fixed the definition of $f^{(l,m)}(\vect s_1)$ by requiring that
$f^{(l,m)}(\vect s_1)$ does \emph{not} contain the term $\propto s_1^{-l-1}$ at $s_1>r_e$ (if it does, one can redefine it by adding a suitable
coefficient times $\phi^{(l,m)}(\vect s_1)$ to the definition of $f^{(l,m)}(\vect s_1)$ to remove such a term),
and similarly fixed the definition of $g^{(l,m)}(\vect s_1)$ by requiring that
it does \emph{not} contain the term $\propto s_1^{-l-1}$ at $s_1>r_e$.
The parameter $a_l$ is the $l$-wave scattering ``length" although its dimension is length raised to the power of $(2l+1)$.
$r_l$ is the $l$-wave effective range, and $r'_l$ is the $l$-wave shape parameter.
The two-body special functions at $s<r_e$ and the parameters $a_l$, $r_l$, $r'_l$ depend on the details of the interaction potential $V_1(s_1)$.
Substituting Eqs.~\eqref{eq:twobodyspecialfunction} into \Eq{eq:wavefuncexpansion} and comparing the result
with \Eq{eq:psiseparate} and \Eq{eq:radialwave}, one can deduce that
$\delta_l(k)$ satisfies the effective range expansion \cite{ERE1, ERE2, ERE3, ERE4}
\begin{equation}\label{eq:effectiverangeexpansion}
k^{2l+1}\cot\delta_l(k)=-a_l^{-1}+r_lk^2/2! + r_l'k^4/4!+O(k^6)
\end{equation}
and $\alpha(l, k)=-a_lk^{l+1}$.

Similarly, one can define the two-body special functions $\tphi^{(l,m)}(\vect s)$, $\tf^{(l,m)}(\vect s)$, and
$\tg^{(l,m)}(\vect s)$ for the collision of particles 1 and 2 which are identical spin-$\downarrow$ fermions. They satisfy
\begin{subequations}
\begin{align}
\widetilde{H}_3\tilde{\phi}^{(l,m)}(\vect s)&=0,\\
\widetilde{H}_3\tilde{f}^{(l,m)}(\vect s)&=\tilde{\phi}^{(l,m)}(\vect s),\\
\widetilde{H}_3\tilde{g}^{(l,m)}(\vect s)&=\tilde{f}^{(l,m)}(\vect s), \dots,
\end{align}
\end{subequations}
where
\beq
\widetilde{H}_3=-\nabla_{\vect s}^2+\frac{2\mu}{\hbar^2}V_3(s).
\eeq
At $s>r_e$ we have the following explicit formulas:
\begin{subequations}\label{eq:twobodyspecialfunctiont}
\begin{align}
    \tphi^\script{l}{m}(\vect s) =& \autobra{\frac{s^l}{(2l+1)!!}-\frac{(2l-1)!!\Tilde{a}_l}{s^{l+1}}}\sqrt{\frac{4\pi}{2l+1}}\sh{l}{m}{s}, \\
    \tf^\script{l}{m}(\vect s) =& -\autobra{\frac{s^{l+2}}{2(2l+3)!!}+\frac{(2l-3)!!\Tilde{a}_l}{2s^{l-1}}+\frac{\Tilde{a}_l\Tilde{r}_ls^l}{2(2l+1)!!}}\nonumber \\
    & \times\sqrt{\frac{4\pi}{2l+1}}\sh{l}{m}{s}, \\
    \tg^\script{l}{m}(\vect s)=& \left[\frac{s^{l+4}}{8(2l+5)!!}-\frac{(2l-5)!!\Tilde{a}_l}{8s^{l-3}}+\frac{\Tilde{a}_l\Tilde{r}_ls^{l+2}}{4(2l+3)!!}\right.\nonumber \\
    & \left.-\frac{\Tilde{a}_l\Tilde{r}_l's^l}{24(2l+1)!!}\right]\sqrt{\frac{4\pi}{2l+1}}\sh{l}{m}{s},
\end{align}
\end{subequations}
where $\Tilde{a}_l$ is the $l$-wave scattering ``length" between two spin-$\downarrow$ fermions, $\Tilde{r}_l$ is the $l$-wave effective range between two spin-$\downarrow$ fermions, and $\Tilde{r}'_l$ is the $l$-wave shape parameter between two spin-$\downarrow$ fermions. For two spin-$\downarrow$ fermions colliding at a small energy
$\hbar^2k^2/m_F$ in the COM frame,
we have the following effective range expansion \cite{ERE1, ERE2, ERE3, ERE4}:
\begin{equation}
k^{2l+1}\cot\delta'_l(k)=-\Tilde{a}_l^{-1}+\Tilde{r}_lk^2/2! + \Tilde{r}'_lk^4/4!+O(k^6),
\end{equation}
where $\delta'_l(k)$ is the $l$-wave scattering phase shift between two spin-$\downarrow$ fermions.

Note that in the two-body special functions $\phi^{(l,m)}(\vect s_1)$, $f^{(l,m)}(\vect s_1)$, and $g^{(l,m)}(\vect s_1)$,
$l$ can be any nonnegative integer.
But in the functions $\tphi^{(l,m)}(\vect s)$, $\tf^{(l,m)}(\vect s)$, and
$\tg^{(l,m)}(\vect s)$, $l$ must be a positive \emph{odd} integer because of Fermi statistics.

\section{\label{app:expansion}DERIVATION OF THE 111-EXPANSION AND THE 21-EXPANSIONS}
We study the expansion of the scattering wave function for three two-component fermions (2 spin-$\downarrow$ fermions and 1 spin-$\uparrow$ fermion). We suppose that the collision energy is zero and the orbital angular momentum $L=1$. The magnetic quantum number $M$ can be $-1$, 0 or 1, and here we derive these expansions for the state of $M=0$.
\begin{figure*}[t]
    \subfigure[Diagram for $t_3^\script{i}{j}$]{\label{fig:tij}
        \includegraphics[width=0.43\linewidth]{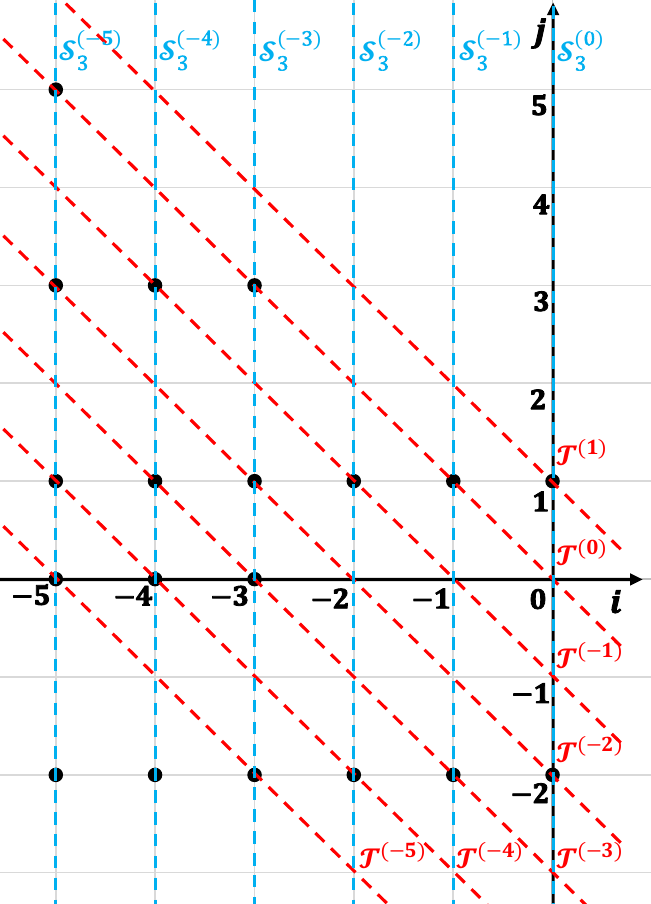}    
    }
    \subfigure[Diagram for $t_1^\script{i}{j}$]{\label{fig:t1ij}
        \includegraphics[width=0.47\linewidth]{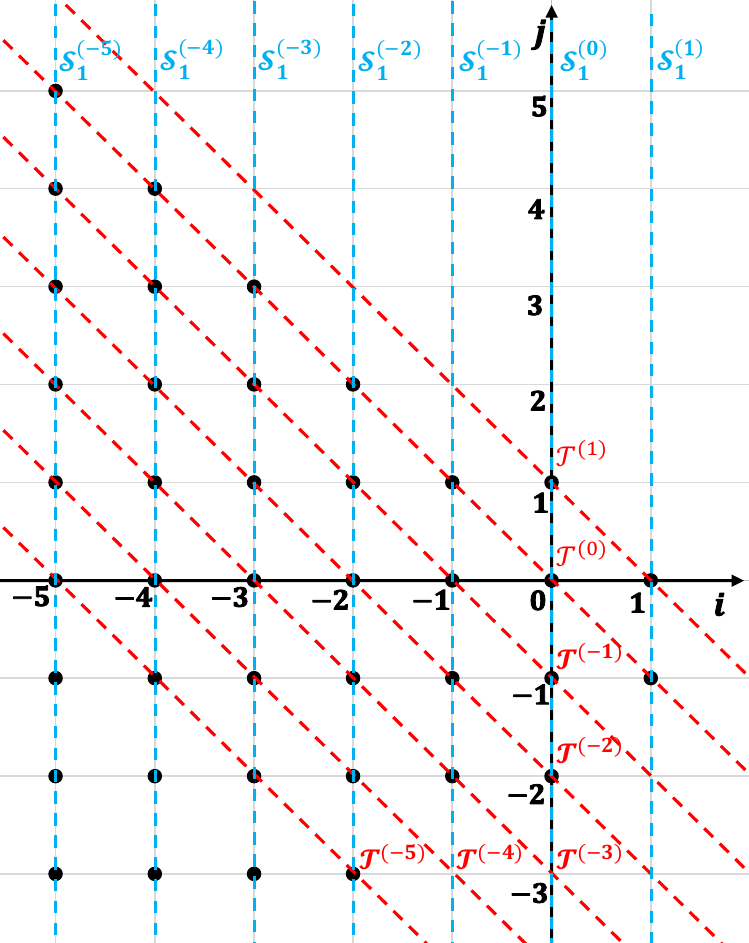}    
    }
    \caption{Diagram of the points representing $t_3^\script{i}{j}$ and $t_1^\script{i}{j}$ on the $(i, j)$ plane. Thick dots represent the points at which $t_3^\script{i}{j}\neq0$ or $t_1^\script{i}{j}\neq0$.}
\end{figure*}

In \Sec{sec:asymptotics} we defined three expansions (\Eq{eq:111def}, \Eq{eq:1st21def} and \Eq{eq:2nd21def}) of the three-body scattering wave function.
The leading order term on the right hand side of \Eq{eq:111def}, $\mt^{(1)}(\vect r_1,\vect r_2,\vect r_3)$, is equal to the $\Psi_0$ defined in \Eq{Psi0}.
The leading order term on the right hand side of \Eq{eq:1st21def}, $\ms_3^{(0)}(\vect R,\vect s)$, scales as $R^0$ because $\Psi_0$ is independent of $R$.
We can use \Eq{eq:sidef}, \Eq{eq:ridef} and \Eq{eq:sdef} to show that $s_z=R_{1z}-\frac12s_{1z}$.
Substituting this into \Eq{Psi0}, we get $\Psi_0=R_{1z}-\frac12s_{1z}$.
At large $R_1$, the leading order term in $\Psi_0$ scales like $R_1^1$,
so we infer that the leading order term on the right hand side of \Eq{eq:2nd21def} is $\msp^{(1)}$.

$\mt^{(-p)}$ satisfies
\begin{equation}\label{eq:laplaceeq}
    -\autoparen{\nabla_1^2+\nabla_2^2+\nabla_3^2}\mt^{(-p)}=0
\end{equation}
when the pairwise distances $s_1$, $s_2$, and $s$ are all nonzero.
When particle 3 is far away from particles 1 and 2, the Schr\"{o}dinger equation \eq{eq:3bodyschrodinger} is simplified as
\begin{equation}
    \autobra{-\frac{\hbar^2}{m_F}\nabla^2_{\vect s}-\frac{3\hbar^2}{4m_F}\nabla^2_{\vect R}+V_3(s)}\Psi=0.
\end{equation}
Thus $\ms_3^{(-q)}$ satisfies
\begin{subequations}\label{eq:hs3}
\begin{align}
    \widetilde{H}_3\ms_3^{(0)} &=\, 0, \label{eq:hs30}\\
    \widetilde{H}_3\ms_3^{(-1)} &=\, 0, \label{eq:hs3-1}\\
    \widetilde{H}_3\ms_3^{(-q)} &=\, \frac{3}{4}\nabla^2_{\vect R}\ms_3^{(-q+2)},~~q\ge2. \label{eq:hs3-2}
\end{align}
\end{subequations}
Similarly, $\msp^{(-q)}$ satisfies
\begin{subequations}\label{eq:hs1}
\begin{align}
    \widetilde{H}_1\msp^{(1)} &=\, 0, \label{eq:hs11}\\
    \widetilde{H}_1\msp^{(0)} &=\, 0, \label{eq:hs10}\\
    \widetilde{H}_1\msp^{(-q)} &=\, \frac{3}{4}\nabla^2_{\vect R_1}\msp^{(-q+2)},~~q\ge1. \label{eq:hs1-1}
\end{align}
\end{subequations}
Furthermore, one can expand $\mt^{(-p)}$ as $\mt^{(-p)}=\sum_i t_3^\script{i}{-p-i}$ when $R\gg s> r_e$ and expand $\mt^{(-p)}$ as $\mt^{(-p)}=\sum_i t_1^\script{i}{-p-i}$ when $R_1\gg s_1> r_e$, where $t_3^\script{i}{j}$ scales as $R^is^j$, and $t_1^\script{i}{j}$ scales as $R_1^is_1^j$. Also, one can expand $\ms_3^{(-q)}$ as $\ms_3^{(-q)}=\sum_j t_3^\script{-q}{j}$ when $R\gg s> r_e$ and expand $\msp^{(-q)}$ as $\msp^{(-q)}=\sum_j t_1^\script{-q}{j}$ when $R_1\gg s_1> r_e$.

The points at which $t_3^\script{i}{j}\neq0$ are shown on the $(i, j)$ plane in Fig. \ref{fig:tij}.
The straight line with slope $-1$ and with vertical intercept $-p$ represents the expansion of $\mt_3^\scriptp{-p}$ and the vertical line $i=-q$ represents the expansion of $\ms_3^\scriptp{-q}$. Similarly, the nonzero $t_1^\script{i}{j}$ terms are shown in Fig. \ref{fig:t1ij}, in which the vertical line $i=-q$ represents the expansion of $\msp^\scriptp{-q}$.


By following the zigzag procedure below, one can determine the 111-expansion \Eq{eq:111expansion} and the 21-expansions \Eq{eq:1st21expansion} and \Eq{eq:2nd21expansion}.

\textit{Step 1.} The leading-order term in the 111-expansion is 
\begin{equation}
    \mt^\scriptp{1} = s_{z} = -\frac{s_{1z}}{2}+R_{1z},
\end{equation}
which indicates that 
\begin{subequations}
\begin{align}
    t_3^\script{0}{1} =&\, s_z, \label{eq:t-01} \\
    t_3^\script{i}{j} =&\, 0, \quad\text{if}\ i+j=1 \ \text{and}\ j\ne1,
\end{align}
\end{subequations}
and
\begin{subequations}
\begin{align}
    t_1^\script{1}{0} =&\, R_{1z}, \label{eq:t1-10}\\
    t_1^\script{0}{1} =&\, -\frac{s_{1z}}{2}, \label{eq:t1-01}\\
    t_1^\script{i}{j} =&\, 0, \quad\text{if}\ i+j=1,\ \text{but}\ i\ne1\ \text{and}\ j\ne1.
\end{align}
\end{subequations}
Since $\mt^\scriptp{2}$, $\mt^\scriptp{3}$, $\mt^\scriptp{4}$, $\dots$ are zero, we get $t_3^\script{i}{j}=t_1^\script{i}{j}=0$ if $i+j\ge2$.

\textit{Step 2a.} Since $\ms_3^\scriptp{0}$ satisfies \Eq{eq:hs30}
and when $s_3>r_e$ we have
\begin{equation}
    \ms_3^\scriptp{0}=t_3^\script{0}{1}+\sum_{j\le0}t_3^\script{0}{j},
\end{equation}
we deduce that 
\begin{equation}\label{eq:s-0}
    \ms_3^\scriptp{0}=\sum_{l,m}c^0_{l,m}\tphi^\script{l}{m}(\vect s).
\end{equation}
Note that $l$ must be equal to 1, because $\tphi^\script{l}{m}(\vect s)$ contains a term proportional to $s^l$ at $s>r_e$. 
Expanding \Eq{eq:s-0} to the order $s^1$ and using \Eq{eq:twobodyspecialfunction}, we find 
\begin{equation}
    t_3^\script{0}{1} = \sum_m c^0_{1,m}\frac{s}{3}\sqrt{\frac{4\pi}{3}}\sh{1}{m}{s}.
\end{equation}
Comparing this result with \Eq{eq:t-01}, we find
\begin{subequations}
\begin{align}
    c^0_{1,0} =&\, 3, \\
    c^0_{1,\pm1} =&\, 0.
\end{align}
\end{subequations}
Thus we have
\begin{equation}
    \ms_3^\scriptp{0} = 3\tphi^\script{1}{0}(\vect s).
\end{equation}
Expanding $\ms_3^\scriptp{0}$ at $s> r_e$, we find
\begin{subequations}
\begin{align}
    t_3^\script{0}{0} =&\, 0, \\
    t_3^\script{0}{-1} =&\, 0, \\
    t_3^\script{0}{-2} =&\, -\frac{3\Tilde{a}_1}{s^3}s_z, \\
    t_3^\script{0}{j} =&\, 0, \quad j\le-3.
\end{align}
\end{subequations}

\textit{Step 2b.} $\msp^\scriptp{1}$ satisfies \Eq{eq:hs11}
and when $s_1>r_e$ we have
\begin{equation}
    \msp^\scriptp{1} = t_1^\script{1}{0} + \sum_{j\le-1}t_1^\script{1}{j}.
\end{equation}
So we deduce that
\begin{equation}\label{eq:s1-1}
    \msp^\scriptp{1} = \sum_{l,m}d^1_{l,m}\phi^\script{l}{m}(\vect s_1),
\end{equation}
where $l$ must be equal to 0 for a reason similar to what we discussed in \textit{Step 2a}. We expand \Eq{eq:s1-1} to the order $s_1^0$ at $s_1> r_e$ and find
\begin{equation}
    t_1^\script{1}{0} = d^1_{0,0}.
\end{equation}
Comparing this result with \Eq{eq:t1-10}, we find
\begin{equation}
    d^1_{0,0} = R_{1z}.
\end{equation}
So we have
\begin{equation}
    \msp^\scriptp{1} = R_{1z}\phi^\script{0}{0}(\vect s_1).
\end{equation}
Expanding $\msp^\scriptp{1}$ at $s_1> r_e$, we find
\begin{subequations}
\begin{align}
    t_1^\script{1}{-1} =&\, -\frac{a_0}{s_1}R_{1z}, \\
    t_1^\script{1}{j} =&\, 0, \quad j\le-2.
\end{align}
\end{subequations}

\textit{Step 3.} At $r_e<s_1\ll R_1$, we expand $\mt^\scriptp{0}$ as
\begin{align}
    \mt^\scriptp{0} =&\, t_1^\script{1}{-1} + t_1^\script{0}{0} + t_1^\script{-1}{1} + t_1^\script{-2}{2} + \dots \nonumber \\
    =&\, t_1^\script{1}{-1} + O(s_1^0).
\end{align}
$-(\nabla^2_{{\vect s}_1}+3\nabla^2_{{\vect R}_1}/4)\mt^\scriptp{0}$ should be a linear combination of the Dirac $\delta$ function of $\vect s_1$
and perhaps some partial derivatives of such a $\delta$ function when $s_1\ll R_1$, since $\mt^\scriptp{0}$ should satisfy the free Schr\"{o}dinger equation outside of the range of the interaction. Note that
\begin{equation}
    -\nabla^2_{{\vect s}_1}t_1^\script{1}{-1} = -4\pi a_0 R_{1z}\delta(\vect s_1),
\end{equation}
so we have
\begin{equation}
    -\autoparen{\nabla^2_{{\vect s}_1}+\frac{3}{4}\nabla^2_{{\vect R}_1}}\mt^\scriptp{0} = -4\pi a_0 R_{1z}\delta(\vect s_1)
\end{equation}
if $\vect s_2\ne0$ and $\vect s\ne0$.
We solve the above equation and find
\begin{equation}
    \mt^\scriptp{0} = -\frac{a_0}{s_1}R_{1z}+\mathcal{T}_2^\scriptp{0},
\end{equation}
where $\mathcal{T}_2^\scriptp{0}$ is a term that satisfies $-\autoparen{\nabla^2_{{\vect s}_1}+\frac{3}{4}\nabla^2_{{\vect R}_1}}\mt_2^\scriptp{0}=0$, possibly except at $\vect s_2=0$ or $\vect s=0$.
At $s\ll R$, we expand $\mathcal{T}^\scriptp{0}$ as
\begin{align}
    \mathcal{T}^\scriptp{0} =&\, t_3^\script{0}{0} + t_3^\script{-1}{1} + t_3^\script{-2}{2} + t_3^\script{-3}{3} + \dots \nonumber \\
    =&\, t_3^\script{0}{0} + O(s^1),
\end{align}
which indicates that $-(\nabla_{\vect s}^2+\frac34\nabla_{\vect R}^2)\mathcal{T}^\scriptp{0}$ vanishes at $\vect s=0$.
So the full $\mt^\scriptp{0}$ (which is anti-symmetric under the interchange of particles 1 and 2) should be
\begin{equation}
    \mt^\scriptp{0} = -\frac{a_0}{s_1}R_{1z} + \frac{a_0}{s_2}R_{2z}.
\end{equation}
If $s\ll R$, we expand $\mt^\scriptp{0}$ as $\sum_{i+j=0}t_3^\script{i}{j}$ and find
\begin{widetext}
\begin{subequations}
\begin{align}
    t_3^\script{-1}{1} =&\, -\frac{a_0}{2}\autoparen{s\frac{R_z R_s}{R^3}+3s_z\frac{1}{R}}, \label{eq:t--11} \\
    t_3^\script{-2}{2} =& 0, \\
    t_3^\script{-3}{3} =&\, \frac{a_0}{16}\autobra{s^3\frac{R_zR_s\autoparen{3R^2-5R_s^2}}{R^7}+3s^2s_z\frac{R^2-3R_s^2}{R^5}}, \\
    t_3^\script{-4}{4} =& 0, \\
    t_3^\script{-5}{5} =&\, -\frac{a_0}{256}\autobra{s^5\frac{R_zR_s\autoparen{15R^4-70R^2R_s^2+63R_s^4}}{R^{11}}+3s^4s_z\frac{3R^4-30R^2R_s^2+35R_s^4}{R^9}},\, \dots,
\end{align}
\end{subequations}
where $R_s\equiv\vect{R}\cdot\hat{\vect s}$. If $s_1\ll R_1$, we expand $\mt^\scriptp{0}$ as $\sum_{i+j=0}t_1^\script{i}{j}$ and find
\begin{subequations}
\begin{align}
    t_1^\script{0}{0} =&\, -\frac{a_0}{2}\frac{R_{1z}}{R_1}, \label{eq:t1-00} \\
    t_1^\script{-1}{1} =&\, \frac{a_0}{4}\autoparen{\spp\frac{\rpz\rps}{\rpp^3}+3\spz\frac{1}{\rpp}}, \\
    t_1^\script{-2}{2} =&\, \frac{a_0}{16}\autobra{\spp^2\frac{\rpz\autoparen{\rpp^2-3\rps^2}}{\rpp^5}-6\spp\spz\frac{\rps}{\rpp^3}}, \\
    t_1^\script{-3}{3} =&\, -\frac{a_0}{32}\autobra{\spp^3\frac{\rpz\rps\autoparen{3\rpp^2-5\rps^2}}{\rpp^7}+3\spp^2\spz\frac{\rpp^2-3\rps^2}{\rpp^5}}, \\
    t_1^\script{-4}{4} =&\, -\frac{a_0}{256}\autobra{\spp^4\frac{\rpz\autoparen{3\rpp^4-30\rpp^2\rps^2+35\rps^4}}{\rpp^9}-12\spp^3\spz\frac{\rps\autoparen{3\rpp^2-5\rps^2}}{\rpp^7}}, \\
    t_1^\script{-5}{5} =&\, \frac{a_0}{512}\autobra{\spp^5\frac{\rpz\rps\autoparen{15\rpp^4-70\rpp^2\rps^2+63\rps^4}}{\rpp^{11}}+3\spp^4\spz\frac{3\rpp^4-30\rpp^2\rps^2+35\rps^4}{\rpp^9}}, \ \dots,
\end{align}
\end{subequations}
\end{widetext}
where $\rps\equiv\vect{R}_1\cdot\vect{\hat{s}}_1$.

\textit{Step 4a.} We expand $\ms_3^\scriptp{-1}$ at $s> r_e$ as
\begin{equation}\label{eq:s--1expand}
    \ms_3^\scriptp{-1} = t_3^\script{-1}{1} + \sum_{j\le0} t_3^\script{-1}{j}.
\end{equation}
$\ms_3^\scriptp{-1}$ satisfies \Eq{eq:hs3-1},
thus we have
\begin{equation}\label{eq:s--1}
    \ms_3^\scriptp{-1} = \sum_{l,m}c^{-1}_{l,m}\tphi^\script{l}{m}(\vect s),
\end{equation}
where $l$ must be equal to 1 because \Eq{eq:s--1} must be compatible with \Eq{eq:s--1expand}. Expanding \Eq{eq:s--1} at $s>r_e$, we get
\begin{equation}
    t_3^\script{-1}{1} = \sum_{m=-1}^1 c^{-1}_{1,m} \frac{s}{3}\sqrt{\frac{4\pi}{3}}\sh{1}{m}{s}.
\end{equation}
Comparing this result with \Eq{eq:t--11}, we find
\begin{subequations}
\begin{align}
    c^{-1}_{1,-1} =&\, -\frac{3\sqrt{2}a_0}{4}\frac{R_z\autoparen{R_x+\I R_y}}{R^3}, \\
    c^{-1}_{1,0} =&\, -\frac{3a_0}{2}\frac{3R^2+R_z^2}{R^3}, \\
    c^{-1}_{1,1} =&\, \frac{3\sqrt{2}a_0}{4}\frac{R_z\autoparen{R_x-\I R_y}}{R^3}.
\end{align}
\end{subequations}
One can re-express these results in terms of the Clebsch-Gordan coefficients and get \Eq{eq:c-11m}. 
Expanding $\ms_3^\scriptp{-1}$ at $s> r_e$, we get
\begin{subequations}
\begin{align}
    t_3^\script{-1}{0} =&\, 0, \\
    t_3^\script{-1}{-1} =&\, 0, \\
    t_3^\script{-1}{-2} =&\, \frac{3a_0\Tilde{a}_1}{2}\autoparen{\frac{1}{s^2}\frac{R_z R_s}{R^3}+3\frac{s_z}{s^3}\frac{1}{R}}, \\
    t_3^\script{-1}{j} =&\, 0, \ j\le-3.
\end{align}
\end{subequations}

\textit{Step 4b.} We expand $\msp^\scriptp{0}$ at $\spp> r_e$ as
\begin{equation}\label{eq:s1-0expand}
    \msp^\scriptp{0} = t_1^\script{0}{1} + t_1^\script{0}{0} + \sum_{j\le-1}t_1^\script{0}{j}.
\end{equation}
$\msp^\scriptp{0}$ satisfies \Eq{eq:hs10},
thus we have
\begin{equation}\label{eq:s1-0}
    \msp^\scriptp{0} = \sum_{l,m}d^0_{l,m}\phi^\script{l}{m}(\vect s_1),
\end{equation}
where $l$ must be 0 or 1, because \Eq{eq:s1-0} must be compatible with \Eq{eq:s1-0expand}. Expanding \Eq{eq:s1-0} at $\spp> r_e$, we have
\begin{align}
    t_1^\script{0}{1} =&\, \sum_{m=-1}^1 d^0_{1,m}\frac{s_1}{3}\sqrt{\frac{4\pi}{3}}\shp{1}{m}{s}, \\
    t_1^\script{0}{0} =&\, d^0_{0,0}.
\end{align}
Comparing these results with \Eq{eq:t1-01} and \Eq{eq:t1-00}, we find
\begin{equation}
    d^0_{0,0} = -\frac{a_0}{2}\frac{\rpz}{\rpp},
\end{equation}
\begin{subequations}
\begin{align}
    d^0_{1,-1} =&\, 0, \\
    d^0_{1,0} =&\, -\frac{3}{2}, \\
    d^0_{1,1} =&\, 0.
\end{align}
\end{subequations}
Expanding $\msp^\scriptp{0}$ at $\spp>r_e$, we find
\begin{subequations}
\begin{align}
    t_1^\script{0}{-1} =&\, \frac{a_0^2}{2}\frac{1}{\spp}\frac{\rpz}{\rpp}, \\
    t_1^\script{0}{-2} =&\, \frac{3a_1}{2}\frac{\spz}{\spp^3}, \\
    t_1^\script{0}{j} =&\, 0, \ j\le-3.
\end{align}
\end{subequations}

One can repeat this procedure step by step, and successively determine $\mt^\scriptp{-1}$, $\ms_3^\scriptp{-2}$, $\msp^\scriptp{-1}$, $\mt^\scriptp{-2}$, $\ms_3^\scriptp{-3}$, $\msp^\scriptp{-2}$, $\dots$, $\mt^\scriptp{-5}$, and $\msp^\scriptp{-5}$. We expand the three-body wave function order by order in this way, and derive the 111-expansion \Eq{eq:111expansion} and the 21-expansions \Eq{eq:1st21expansion} and \Eq{eq:2nd21expansion}.

The following is the list of the coefficients $c^{-i}_{l,m}$:

\begin{align}
    c^{-3}_{1,m} =&\, \autoparen{\frac{2\sqrt{2\pi}J_{3}}{R^3}-\frac{9\sqrt{\pi}a_0\Tilde{a}_1\Tilde{r}_1}{2\sqrt{2}R^3}}\cg{1}{0}{1}{m}{2}{-m}\sh{2}{-m}{R} \nonumber \\
    &-\kroneckerdelta{m}{0}2\sqrt{\pi}\frac{J_{3}+3K_{3}}{R^3}\cg{1}{0}{1}{0}{0}{0}\sh{0}{0}{R}, \label{eq:c-start}\\
    J_{3} =&\, -\frac{9a_1}{2}+\frac{2\omega a^3_0}{\pi}\autoparen{16-3\sqrt{3}\pi}, \\
    K_{3} =&\, \frac{3a_1}{2}-\frac{2\omega a_0^3}{\pi}\autoparen{6-\sqrt{3}\pi},\\
    c^{-3}_{3,m} =&\, \frac{5\sqrt{21\pi}a_0}{2R^3}\cg{1}{0}{3}{m}{4}{-m}\sh{4}{-m}{R}\nonumber\\
    &-\frac{39\sqrt{7\pi}a_0}{2R^3}\cg{1}{0}{3}{m}{2}{-m}\sh{2}{-m}{R}, 
\end{align}
\begin{align}
     c^{-4}_{1,m} =&\, -\frac{2\sqrt{2}}{3\sqrt{\pi}R^4}\autoparen{3\sqrt{3}\xi_4+8\pi\zeta_4}\cg{1}{0}{1}{m}{2}{-m}\sh{2}{-m}{R} \nonumber \\
    & +\kroneckerdelta{m}{0}\frac{1}{2\sqrt{\pi}R^4}\autoparen{\sqrt{3}\xi'_4+\pi\zeta'_4}\cg{1}{0}{1}{0}{0}{0}\sh{0}{0}{R},\\
    \xi_4 =&\, 6a_0\Tilde{a}_1+57a_0 a_1+20a_0\omega_3-15a_0^2\Tilde{a}_1\Tilde{r}_1, \\
    \zeta_4 =&\, -15a_0a_1-4a_0\omega_3+3a_0^2\Tilde{a}_1\Tilde{r}_1, \\
    \xi'_4 =&\, 8a_0\omega_3-42a_0a_1-30a_0\Tilde{a}_1+3a_0^2\Tilde{a}_1\Tilde{r}_1, \\
    \zeta'_4 =&\, 2\autoparen{8a_0\omega_3+30a_0a_1+3a_0^2\tilde{a}_1\tilde{r}_1}/3, \\
    c^{-4}_{3,m} =&\, -\frac{16\sqrt{21\pi}\omega'''a_0^2}{3\pi R^4}\cg{1}{0}{3}{m}{4}{-m}\sh{4}{-m}{R} \nonumber \\
    &+\frac{8\sqrt{7\pi}a_0^2}{R^4}\cg{1}{0}{3}{m}{2}{-m}\sh{2}{-m}{R}, \\
    \omega''' =&\, 189\sqrt{3}-104\pi, 
\end{align}
\begin{align}
    c^{-5}_{1,m} =&\, -\frac{\sqrt{\pi}\omega_5}{R^5}\cg{1}{0}{1}{m}{2}{-m}\sh{2}{-m}{R} \nonumber \\
    &+\kroneckerdelta{m}{0}\frac{\sqrt{\pi}\omega'_5}{R^5}\cg{1}{0}{1}{0}{0}{0}\sh{0}{0}{R},\\
    \omega_5 =&\, \frac{\sqrt{2}a_0^2a_1}{2\pi}\autoparen{39\sqrt{3}-10\pi}-\sqrt{2}a_0\omega_4, \\
    \omega'_5 =&\, \frac{a_0^2a_1}{\pi}\autoparen{87\sqrt{3}-50\pi}-10a_0\omega_4+\frac{18}{\pi}\omega a_0^3\Tilde{a}_1\Tilde{r}_1,\\
    c^{-5}_{3,m} =&\, -\frac{\sqrt{21\pi}}{8R^5}\Big[420a_1+105a_0\Tilde{a}_3\Tilde{r}_3\nonumber \\
    &+16\autoparen{315\sqrt{3}-1712/\pi}\omega a_0^3\Big]\cg{1}{0}{3}{m}{4}{-m}\sh{4}{-m}{R} \nonumber \\
    &+\frac{30\sqrt{7}\omega a_0^3}{\sqrt{\pi}R^5}\cg{1}{0}{3}{m}{2}{-m}\sh{2}{-m}{R}, \\
    c^{-5}_{5,m} =&\, \frac{945\sqrt{11\pi}a_0}{8\sqrt{2}R^5}\cg{1}{0}{5}{m}{6}{-m}\sh{6}{-m}{R}\nonumber \\
    &-\frac{735\sqrt{165\pi}a_0}{8R^5}\cg{1}{0}{5}{m}{4}{-m}\sh{4}{-m}{R}. \label{eq:c-end}
\end{align}

The following is the list of the coefficients $d^{-i}_{l,m}$:
\begin{align}
    d^{-2}_{0,0} =&\, -\frac{2\omega_3\sqrt{\pi}}{\sqrt{3}R_1^2}\cg{1}{0}{0}{0}{1}{0}\shp{1}{0}{R}, \label{eq:d-start} \\
    d^{-2}_{1,m} =&\, -\frac{2\sqrt{2}\omega' a_0^2}{3\sqrt{\pi}R_1^2}\cg{1}{0}{1}{m}{2}{-m}\shp{2}{-m}{R} \nonumber \\
    &-\kroneckerdelta{m}{0}\frac{\omega'' a_0^2}{3\sqrt{\pi}R_1^2}\cg{1}{0}{1}{0}{0}{0}\shp{0}{0}{R}, \\
    d^{-2}_{2,m} =&\, -\frac{3\sqrt{5\pi}a_0}{4R_1^2}\cg{1}{0}{2}{m}{3}{-m}\shp{3}{-m}{R} \nonumber \\
    &+\frac{3\sqrt{30\pi}a_0}{2R_1^2}\cg{1}{0}{2}{m}{1}{-m}\shp{1}{-m}{R}, 
\end{align}
\begin{align}
    d^{-3}_{0,0} =&\, \frac{2\sqrt{\pi}\omega_4}{\sqrt{3}R_1^3}\cg{1}{0}{0}{0}{1}{0}\shp{1}{0}{R}, \\
    d^{-3}_{1,m} =&\, \frac{\ell_1}{4\sqrt{2\pi}R_1^3}\cg{1}{0}{1}{m}{2}{-m}\shp{2}{-m}{R}\nonumber \\ 
    &-\kroneckerdelta{m}{0}\frac{4\omega a_0^3}{\sqrt{\pi}R_1^3}\cg{1}{0}{1}{0}{0}{0}\shp{0}{0}{R}, \\
    d^{-3}_{2,m} =&\, \frac{8\sqrt{5}a_0^2}{\sqrt{\pi}R_1^3}\autoparen{-9\sqrt{3}+5\pi}\cg{1}{0}{2}{m}{3}{-m}\shp{3}{-m}{R} \nonumber \\
    &-\frac{3\sqrt{10}a^2_0}{\sqrt{\pi}R_1^3}\cg{1}{0}{2}{m}{1}{-m}\shp{1}{-m}{R}, \\
    d^{-3}_{3,m} =&\, -\frac{5\sqrt{21\pi}a_0}{4R_1^3}\cg{1}{0}{3}{m}{4}{-m}\shp{4}{-m}{R}\nonumber \\
    &+\frac{39\sqrt{7\pi}a_0}{4R_1^3}\cg{1}{0}{3}{m}{2}{-m}\shp{2}{-m}{R}, 
\end{align}
\begin{align}
    d^{-4}_{0,0} =&\, \frac{2\sqrt{\pi}\Omega_5}{\sqrt{3}R_1^4}\cg{1}{0}{0}{0}{1}{0}\shp{1}{0}{R}, \\
    d^{-4}_{1,m} =&\, \autoparen{\frac{\sqrt{2}\omega' a_0^2a_1r_1}{\sqrt{\pi}R_1^4}-\frac{2\sqrt{2}\Omega_4}{3\sqrt{\pi}R_1^4}}\cg{1}{0}{1}{m}{2}{-m}\shp{2}{-m}{R} \nonumber \\
    &+\kroneckerdelta{m}{0}\frac{a_0\Omega'_3}{12\sqrt{\pi}R_1^4}\cg{1}{0}{1}{0}{0}{0}\shp{0}{0}{R} \\
    \Omega_4 =&\, 2\omega'\omega_3a_0+3\autoparen{-57\sqrt{3}+40\pi} a_0\Tilde{a}_1 \nonumber \\
    &-\frac{3}{2}\autoparen{-51\sqrt{3}+40\pi}a_0a_1, \\
    \Omega'_3 =&\, a_1\autobra{60\pi-36\sqrt{3}-a_0r_1\autoparen{9\sqrt{3}+6\pi}}\nonumber \\
    & +4\autobra{\Tilde{a}_1\autoparen{63\sqrt{3}-30\pi}-\omega_3\autoparen{6\sqrt{3}+4\pi}}, \\
    d^{-4}_{2,m} =&\, \frac{\sqrt{5\pi}}{2R_1^4}\left[\frac{6\omega a_0^3}{\sqrt{3}\pi}\autoparen{82\sqrt{3}-45\pi}\right.\nonumber\\
    & \left.-45\autoparen{\Tilde{a}_1+\frac{a_1}{2}}+\frac{45}{8}a_0a_2r_2\right]\cg{1}{0}{2}{m}{3}{-m}\shp{3}{-m}{R} \nonumber \\
    &-\frac{2\sqrt{30}\omega a_0^3}{\sqrt{\pi}R_1^4}\cg{1}{0}{2}{m}{1}{-m}\shp{1}{-m}{R}, \\
    d^{-4}_{3,m} =&\, \frac{8\sqrt{21}\omega''' a_0^2}{3\sqrt{\pi}R_1^4}\cg{1}{0}{3}{m}{4}{-m}\shp{4}{-m}{R} \nonumber \\
    &-\frac{4\sqrt{7\pi}a_0^2}{R_1^4}\cg{1}{0}{3}{m}{2}{-m}\shp{2}{-m}{R},     \\
    d^{-4}_{4,m} =&\, -\frac{105\sqrt{15\pi}a_0}{16R_1^4}\cg{1}{0}{4}{m}{5}{-m}\shp{5}{-m}{R} \nonumber \\
    &+\frac{255\sqrt{3\pi}a_0}{4R_1^4}\cg{1}{0}{4}{m}{3}{-m}\shp{3}{-m}{R}, 
\end{align}
\begin{align}
    d^{-5}_{0,0} =&\, \frac{a_0r'_0}{24}\Tilde{\Tilde{d}}^{-1}_{0,0} + \frac{a_0r_0}{2}\Tilde{d}^{-3}_{0,0} \nonumber \\
    &+ \frac{R_{1z}}{R_1^6}\left[\frac{135\sqrt{3}a_0a_2}{4\pi}-\frac{3\sqrt{3}D}{8\pi^3}\right.\nonumber \\
    &\left.+\frac{2a_0\Omega_5}{3\pi}\autoparen{\sqrt{3}+\pi+6\sqrt{3}\ln\frac{b}{R_1}}-\Omega_6\right], \label{eq:d-500} \\
    \Omega_6 =&\, \frac{a_1}{2\pi^2}\left[\autoparen{2\pi/3-5\sqrt{3}/4}\ell_1-8\omega a_0^3\left(4\sqrt{3}\right.\right.\nonumber \\
    &\left.\left.-3\sqrt{3}\gamma+\pi/6+3\sqrt{3}\ln\frac{b}{R_1}\right)\right]\nonumber \\
    &+\frac{\Tilde{a}_1}{\pi^2}\left[\autoparen{2\pi/3-5\sqrt{3}/4}\ell_3-8\omega a_0^3\left(4\sqrt{3}\right.\right.\nonumber \\
    &\left.\left.-3\sqrt{3}\gamma+\pi/6+3\sqrt{3}\ln\frac{b}{R_1}\right)\right], \\
    d^{-5}_{1,m} =&\, \frac{\sqrt{2\pi}}{4R_1^5}\Big[a_0^2\autoparen{2\Tilde{a}_1-a_1}\autoparen{39\sqrt{3}-10\pi}/\pi \nonumber \\
    & - 2a_0\omega_4\Big]\cg{1}{0}{1}{m}{2}{-m}\shp{2}{-m}{R} \nonumber \\
    &-\delta_{m,0}\frac{\sqrt{\pi}}{4R_1^5}\Big[ a_0^2\autoparen{2\Tilde{a}_1-a_1}\autoparen{174\sqrt{3}-100\pi}/\pi \nonumber\\
    &+36\omega a_0^3a_1r_1/\pi - 20a_0\omega_4\Big]\cg{1}{0}{1}{0}{0}{0}\shp{0}{0}{R}, \\
    d^{-5}_{2,m} =&\, \frac{10a_0 \widetilde{\Omega}_3}{\sqrt{15\pi}R_1^5}\cg{1}{0}{2}{m}{3}{-m}\shp{3}{-m}{R} \nonumber \\
    &- \frac{\sqrt{10}a_0\widetilde{\Omega}'_3}{\sqrt{\pi}R_1^5}\cg{1}{0}{2}{m}{1}{-m}\shp{1}{-m}{R}, \\
    \widetilde{\Omega}_3 =&\, 27\autoparen{9a_0a_2r_2-8\omega_3+84\Tilde{a}_1+42a_1} \nonumber \\
    &-\sqrt{3}\pi\autoparen{45a_0a_2r_2-40\omega_3+396\Tilde{a}_1+198a_1}, \\
    \widetilde{\Omega}'_3 =&\, 6\omega_3+9a_0a_2r_2/2 \nonumber \\
    & -\sqrt{3}\autoparen{\Tilde{a}_1+a_1/2}\autoparen{21\sqrt{3}-16\pi}, \\
    d^{-5}_{3,m} =&\, -\frac{\sqrt{21}}{16\sqrt{\pi}R_1^5}\Big[16\omega a_0^3\autoparen{1712-315\sqrt{3}\pi} \nonumber \\
    &-105\pi\autoparen{a_0a_3r_3+8\Tilde{a}_1-4a_1}\Big]\cg{1}{0}{3}{m}{4}{-m}\shp{4}{-m}{R} \nonumber \\
    &- \frac{15\sqrt{7}\omega a_0^3}{\sqrt{\pi}R_1^5}\cg{1}{0}{3}{m}{2}{-m}\shp{2}{-m}{R}, \\ 
    d^{-5}_{4,m} =&\, -\frac{8\sqrt{15}a_0^2}{\sqrt{\pi}R_1^5}\autoparen{1269\sqrt{3}-700\pi}\cg{1}{0}{4}{m}{5}{-m}\shp{5}{-m}{R} \nonumber \\
    &-\frac{4\sqrt{3}a_0^2}{\sqrt{\pi} R_1^5}\autoparen{27\sqrt{3}-8\pi}\cg{1}{0}{4}{m}{3}{-m}\shp{3}{-m}{R}, \\
    d^{-5}_{5,m} =&\, -\frac{945\sqrt{11\pi}a_0}{16\sqrt{2}R_1^5}\cg{1}{0}{5}{m}{6}{-m}\shp{6}{-m}{R} \nonumber \\
    &+\frac{735\sqrt{165\pi}a_0}{16R_1^5}\cg{1}{0}{5}{m}{4}{-m}\shp{4}{-m}{R}. \label{eq:d-end}
\end{align}

One can generate the expansions of the three-body scattering wave functions at state $\ket{L=1, M=-1}$ or $\ket{L=1, M=1}$ by applying the ladder operator $J_{\pm}$ on the expansions of state $\ket{L=1, M=0}$. 
Here $J_{\pm}=J_x\pm\I J_y$, and $J_x$ and $J_y$ are the projections of the total orbital angular momentum operator in the $x$ and the $y$ directions, respectively. We have
\begin{align}
    J_x =&\, \I\hbar\left(\sin\phi_{s_i}\frac{\partial}{\partial\theta_{s_i}}+\cot\theta_{s_i}\cos\phi_{s_i}\frac{\partial}{\partial\phi_{s_i}}\right. \nonumber \\
    & \left.+\sin\phi_{R_i}\frac{\partial}{\partial\theta_{R_i}}+\cot\theta_{R_i}\cos\phi_{R_i}\frac{\partial}{\partial\phi_{R_i}}\right), \\
    J_y =&\, \I\hbar\left(-\cos\phi_{s_i}\frac{\partial}{\partial\theta_{s_i}}+\cot\theta_{s_i}\sin\phi_{s_i}\frac{\partial}{\partial\phi_{s_i}}\right. \nonumber \\
    & \left.-\cos\phi_{R_i}\frac{\partial}{\partial\theta_{R_i}}+\cot\theta_{R_i}\sin\phi_{R_i}\frac{\partial}{\partial\phi_{R_i}}\right), 
\end{align}
where $\theta_{s_i}$, $\phi_{s_i}$, $\theta_{R_i}$, and $\phi_{R_i}$ are spherical coordinates such that
\begin{subequations}
\begin{align}
    s_{ix}&=s_i\sin\theta_{s_i}\cos\phi_{s_i}, \\
    s_{iy}&=s_i\sin\theta_{s_i}\sin\phi_{s_i}, \\
    s_{iz}&=s_i\cos\theta_{s_i}, \\
    R_{ix}&=R_i\sin\theta_{R_i}\cos\phi_{R_i}, \\
    R_{iy}&=R_i\sin\theta_{R_i}\sin\phi_{R_i}, \\
    R_{iz}&=R_i\cos\theta_{R_i}.
\end{align}
\end{subequations}

One can also study those $\Psi$'s with leading order term $\Psi_0$'s scaling as $B^{p'}$, where $p'>p_\text{min}$. For example, if we set $p'=2$, then there are \textit{nine} linearly independent leading order terms, one of which has $L=0$ and $M=0$, i.e., $\vect s\cdot\vect R$, three of which have $L=1$  and $M=0, \pm1$, i.e., $(\vect s\cross\vect{R})_z$ and $\mp[(\vect s\cross\vect{R})_x\pm\I (\vect s\cross\vect R)_y]/\sqrt{2}$,
and five of which have $L=2$ and $M=0, \pm1, \pm2$, i.e., $(s_x+\I s_y)(R_x+\I R_y)$, $(s_x+\I s_y)R_z+s_z(R_x+\I R_y)$, $-\vect s\cdot\vect R+3s_zR_z$, $(s_x-\I s_y)R_z+s_z(R_x-\I R_y)$, and $(s_x-\I s_y)(R_x-\I R_y)$. These wave functions are usually less important than the wave functions with $p=1$ for dilute ultracold gases, and will not be studied in this paper.

\begin{widetext}
\section{\label{app:born_approximation}THE BORN APPROXIMATION FOR THE THREE-BODY WAVE FUNCTION}
For weak interactions, we expand the three-body wave function as a Born series:
\begin{equation}
    \Psi = \Psi_0 + \widehat{G}\mathcal{V}\Psi_0 + \autoparen{\widehat{G}\mathcal{V}}^2\Psi_0 + \dots,
\end{equation}
where $\mathcal{V}=U(s_1,s_2,s_3)+\sum_iV_i(s_i)$ is the total interaction operator, $\widehat{G}=-\widehat{H}_0^{-1}$ is the Green's operator, $\widehat{H}_0$ is the three-body kinetic energy operator, and $\Psi_0=s_z$ is the wave function of three free fermions. 
We assume that $V_i(s_i)$ vanishes if $s_i>r_e$ and that $U(s_1,s_2,s_3)$ vanishes if $s_1>r_e$ or $s_2>r_e$ or $s_3>r_e$.

For the first order term in the Born series, we have
\begin{align}
    \widehat{G}\mathcal{V}\Psi_0 =&\, \widehat{G}U\Psi_0 + \sum_{i=1}^3\widehat{G}V_i\Psi_0, \nonumber \\
    =&\, \frac{m_F}{\hbar^2}\int \diff^6\rho'_i\,\mg\autoparen{\brho_i-\brho'_i}U(s'_1,s'_2,s'_3)\Psi_0\autoparen{\brho'_i} +\frac{m_F}{\hbar^2}\sum_{i=1}^3\int \diff^6\rho'_i\,\mg\autoparen{\brho_i-\brho'_i} V_i(s'_i)\Psi_0\autoparen{\brho'_i},
\end{align}
where $\brho_i=(\vect s_i, 2\vect R_i/\sqrt{3})$ and $\brho'_i=(\vect s'_i, 2\vect R'_i/\sqrt{3})$ are six dimensional vectors, and $\mg\autoparen{\brho_i-\brho'_i}$ is the Green's function in six dimensional space,
\begin{equation}
    \mg\autoparen{\brho_i-\brho'_i} = -\frac{1}{4\pi^3|\brho_i-\brho'_i|^4}.
\end{equation}
Defining $\Psi_{1,i}=\widehat{G}V_i\Psi_0$, we have
\begin{align}
    \Psi_{1,i} =&\, -\frac{m_F}{\hbar^2}\int\diff^6\rho'_i \frac{1}{4\pi^3|\brho_i-\brho'_i|^4}V_i(s'_i)\Psi_0(\brho'_i), \nonumber \\
    =&\, -\frac{\sqrt{3}m_F}{8\pi^3\hbar^2}\int\diff^3s'_i\int\diff^3R'_i\frac{V_i(s'_i)\Psi_0}{\autobra{\frac{3}{4}(\vect s_i-\vect s'_i)^2+(\vect R_i-\vect R'_i)^2}^2}.
\end{align}
Carrying out this integral, we get
\begin{subequations}
\begin{align}\label{eq:psi1-2born}
    \Psi_{1,1} =&\, \frac{1}{6}s_{1z}\autobra{\frac{\alpha_{3,1}(s_1)}{s_1^3}+\Bar{\alpha}_{0,1}(s_1)} - R_{1z}\autobra{\frac{\alpha_{1,1}(s_1)}{s_1}+\Bar{\alpha}_{0,1}(s_1)}, \\
    \Psi_{1,2} =&\, \frac{1}{6}s_{2z}\autobra{\frac{\alpha_{3,2}(s_2)}{s_2^3}+\Bar{\alpha}_{0,2}(s_2)} - R_{2z}\autobra{\frac{\alpha_{1,2}(s_2)}{s_2}+\Bar{\alpha}_{0,2}(s_2)}, \\
    \Psi_{1,3} =&\, -\frac{1}{3}s_z\autobra{\frac{\alpha_{3,3}(s)}{s^3}+\Bar{\alpha}_{0,3}(s)},
\end{align}
\end{subequations}
where
\begin{subequations}
\begin{align}
    \alpha_{n,i}(s_i) \equiv&\, \frac{m_F}{\hbar^2}\int_0^{s_i} \diff s'\,s'^{n+1}V_i(s'), \\
    \Bar{\alpha}_{n,i}(s_i) \equiv&\, \frac{m_F}{\hbar^2}\int_{s_i}^\infty \diff s'\,s'^{n+1}V_i(s').
\end{align}
\end{subequations}
At $s_i>r_e$,
\begin{align}
\alpha_{n,i}(s_i)=&\alpha_{n,i}\equiv\frac{m_F}{\hbar^2}\int_0^{\infty} \diff s'\,s'^{n+1}V_i(s'),\\
\Bar{\alpha}_{n,i}(s_i)=&0.
\end{align}
If $s_1$, $s_2$, and $s_3$ are all greater than $r_e$,
\begin{equation}
    \sum_{i=1}^3\Psi_{1,i} = -\frac{1}{3}s_z\frac{\alpha_{3,3}}{s^3} +\frac{1}{6}s_{1z}\frac{\alpha_{3,1}}{s_1^3} - R_{1z}\frac{\alpha_{1,1}}{s_1} + \frac{1}{6}s_{2z}\frac{\alpha_{3,2}}{s_2^3} + R_{2z}\frac{\alpha_{1,2}}{s_2}.
\end{equation}

Defining $\Psi_{1,U}=\widehat{G}U\Psi_0$, we have
\begin{align}\label{eq:psi1-3}
    \Psi_{1,U} =&\, -\frac{m_F}{\hbar^2}\int\diff^6\rho'\frac{1}{4\pi^3|\brho-\brho'|^4}U(\rho')\Psi_0(\rho') \nonumber \\
    =&\, -\frac{\sqrt{3}m_F}{8\pi^3\hbar^2}\int\diff^3s'\int\diff^3R' \frac{U(s'_1,s'_2,s')}{\autobra{\frac{3}{4}(\vect s-\vect s')^2+(\vect R-\vect R')^2}^2}s'_z.
\end{align}
Since the potential $U(s_1,s_2,s_3)$ is finite-ranged, the factor $\frac{1}{\autobra{\frac{3}{4}(\vect s-\vect s')^2+(\vect R-\vect R')^2}^2}$ in the integrand
in \Eq{eq:psi1-3} can be expanded when $s$ and $R$ go to infinity at a fixed ratio $s/R$:
\begin{align}
    \frac{1}{\autobra{\frac{3}{4}(\vect s-\vect s')^2+(\vect R-\vect R')^2}^2} = \frac{1}{B^4} + \frac{3\vect s\cdot\vect s'+4\vect R\cdot\vect R'}{B^6} + O(B^{-6}).
\end{align}
By symmetry, only the term $3\vect{s}\cdot\vect{s}'/B^6$ contributes to the integral. Thus we get
\begin{align}
    \Psi_{1,U} =&\, -\frac{3\sqrt{3}m_F}{8\pi^3\hbar^2}\frac{s}{B^6}\int_0^\infty\diff s'\int_0^\infty\diff R'\int_0^\pi\diff\theta' s'^4R'^2\sin\theta'U(\vect s',\vect R')\times\int_{-\pi}^\pi\diff\alpha'\int_{-\pi}^\pi\diff\gamma'\int_{0}^\pi\diff\beta'\sin\beta'\cos\beta'\hat{s}'_z + O(B^{-6})\nonumber \\
    =&\, -\frac{\sqrt{3}}{\pi}\frac{s_z}{B^6}\Lambda + O(B^{-6}),
\end{align}
where $\alpha'$, $\beta'$ and $\gamma'$ are three Euler's angles, $\theta'$ is the angle between $\vect s'$ and $\vect R'$, and
\begin{align}
    \hat{s}'_z = \hat{s}'\cdot\hat{z} = \sin\theta_z\cos\phi_z\sin\beta'\sin\alpha'-\sin\theta_z\sin\phi_z\sin\beta'\cos\alpha'+\cos\theta_z\cos\beta',
\end{align}
and
\begin{align}
    \Lambda = \frac{m_F}{\hbar^2}\int_0^\infty\diff s'\, s'^4\int_0^\infty\diff R'\, R'^2 \int_0^\pi\diff\theta'\sin\theta'\, U(\vect s', \vect R').
\end{align}

\begin{acknowledgments}
This work was supported by the National Natural Science Foundation of China Grant No. 92365202
and the National Key R$\&$D Program of China (Grants No. 2019YFA0308403 and No. 2021YFA1400902).
We thank Junjie Liang for discussions.
\end{acknowledgments}
\end{widetext}

\nocite{*}

\bibliography{Three-body_scattering_hypervolume_of_two-component_fermions_in_three_dimensions}

\end{document}